\documentclass{PoS}
\usepackage{amsmath,mathrsfs,fixmath}
\def\Tr{{\rm Tr}}

\newcommand{\dgamma}{\hat{\gamma}}

\title{Form factors for rare $\mathbold{B}$ decays: strategy, methodology,
and numerical study }

\ShortTitle{Form factors for rare $B$ decays}

\author{\speaker{Zhaofeng Liu}$\:\:^a$, Stefan Meinel$\:^a$, Alistair Hart$\:^b$, Ron R. Horgan$\:^a$,
Eike H. M\"uller$\:^b$, Matthew Wingate\thanks{Poster presenter.}$\:\:\:^a$  \\  \\
\llap{$^a$}DAMTP, University of Cambridge, Wilberforce Road, Cambridge CB3 0WA, UK\\
\llap{$^b$}SUPA, School of Physics and Astronomy, University of Edinburgh, Edinburgh, UK\\
\\  \\
E-mail: \email{Z.Liu@damtp.cam.ac.uk} \\
\phantom{E-mail: }\email{S.Meinel@damtp.cam.ac.uk} \\
\phantom{E-mail: }\email{M.Wingate@damtp.cam.ac.uk}
}

\abstract{

  We investigate the combined use of moving NRQCD and stochastic sources in
  lattice calculations of form factors describing rare $B$ and $B_s$ decays.
  Moving NRQCD leads to a reduction of discretisation errors compared to standard
  NRQCD. Stochastic sources are tested for reduction of statistical errors.

}

\FullConference{The XXVII International Symposium on Lattice Field Theory\\
                July 26-31, 2009\\
                Peking University, Beijing, China}

\begin{document}

\section{Introduction} 

Rare $B$ decays proceeding through $b\rightarrow s$ flavour changing neutral current transitions
provide particularly sensitive probes for physics beyond the Standard Model.
Experimental uncertainties in $B\rightarrow K^*\gamma$ have been reduced to the
few-percent level, and those for $B\to K^{(*)}\ell\ell$ are presently around
10 percent~\cite{Barberio:2008fa}. So far, measurements are consistent
with Standard Model expectations \cite{Ball:2006eu,Ali:1999mm}.
Future experiments like LHCb are going to reduce the experimental
uncertainties further. Thus, we are entering an era of precision flavour
physics, requiring experimentalists and theorists to focus on reducing
statistical errors and employing multiple cross-checks to further
quantify systematic uncertainties.

Lattice QCD can contribute to this by providing first-principles nonperturbative
calculations of hadronic matrix elements relevant for rare $B$ decays.
In this report, we present our current progress in doing these computations.
We use a non-relativistic effective action for the $b$ quark
called moving NRQCD (mNRQCD) \cite{Horgan:2009ti}, which helps reduce
discretisation errors in the final state meson at large recoil. This is
achieved by giving the $B$ meson a significant velocity in the lattice frame,
so that at a given $q^2$ the momentum of the final meson is smaller.

Previous progress of this project was reported in ~\cite{Meinel:2008th}.
In this work, we investigate how far statistical errors can be reduced with
the help of stochastic (random wall) sources, and examine different fitting strategies.
Furthermore, all calculations are now done with the full $\mathcal{O}(1/m_b^2)$
\mbox{mNRQCD} action exactly as in \cite{Horgan:2009ti}.

This report is organised as follows. We discuss the phenomenology of rare $B$ decays
and our strategy for the lattice calculation in Sec.~\ref{sec:strategy}. Details of the operator
matching for heavy-light currents with mNRQCD are given in Sec.~\ref{sec:current_matching}. Then,
in Sec.~\ref{sec:stochstic_sources} we explain the use of stochastic sources and compare their effectiveness
to point sources using numerical calculations.
Finally we show some preliminary results for the form factors in Sec~\ref{sec:FF_results}
and conclude in Sec.~\ref{sec:discussion}.

\section{Physics of rare $B$ decays and strategy for the lattice calculation}
\label{sec:strategy}

\begin{figure}[t!]
\null
\vspace{-1ex}
\begin{center}
\includegraphics[clip=true,width=0.25\textwidth]{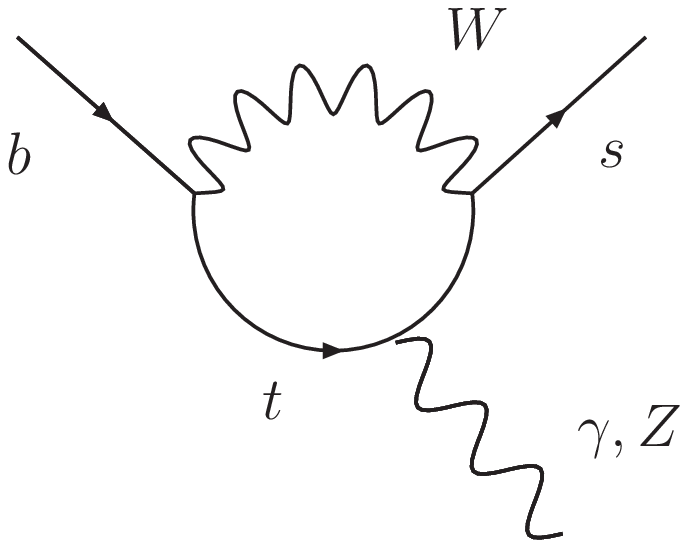}~~~~~
\includegraphics[clip=true,width=0.25\textwidth]{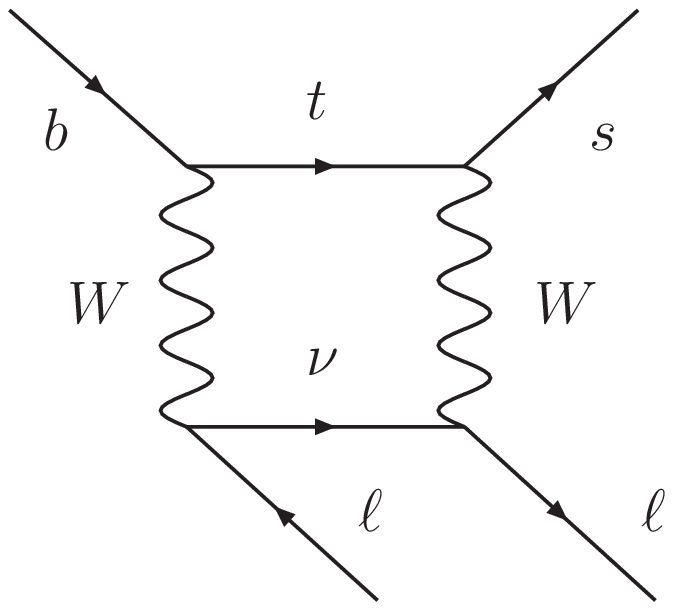}
\end{center}
\vspace{-4ex}
\caption{\label{fig:PengBox}Penguin and box diagrams governing
$b\to s\gamma$ and $b\to s\ell\ell$ in the Standard Model.  (Decay of
photon or $Z$ to $\ell\ell $ not shown in the penguin diagram.)}
\end{figure}

\begin{figure}[t!]
\null
\vspace{-1ex}
\begin{center}
\includegraphics[clip=true,width=0.25\textwidth]{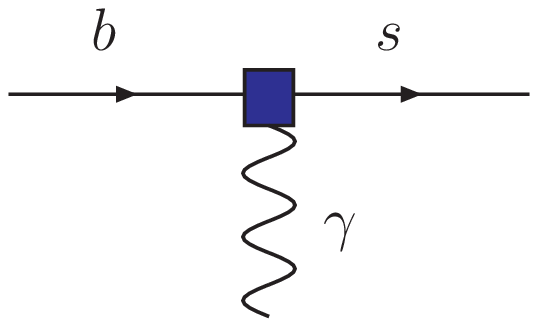}~~~~~
\includegraphics[clip=true,width=0.25\textwidth]{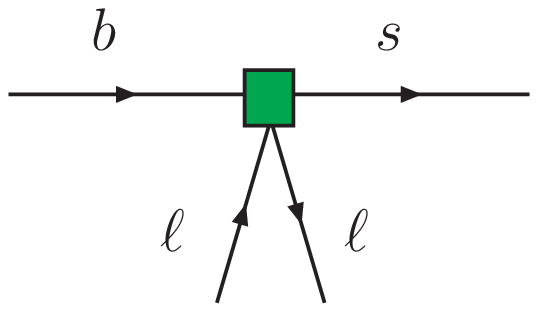}
\end{center}
\vspace{-4ex}
\caption{\label{fig:2quark}Decays $b\to s\gamma$ and $b\to s\ell\ell$
via short distance operators in the effective weak theory.}
\end{figure}

The starting point for studying weak decays of hadrons is the effective
weak Hamiltonian. For $b\rightarrow s$ transitions in the Standard Model, 
the governing Hamiltonian is\footnote{To obtain Eq.(\ref{eq:h_eff}),
$V_{ub}V_{us}^*\ll V_{tb}V_{ts}^*$ and CKM unitarity are used.}
\begin{equation}
\mathcal{H}_{\mathrm{eff}}^{b\to s} ~=~ 
-\frac{G_F}{\sqrt{2}} \,V_{ts}^* V_{tb} \sum_i C_i(\mu) \,Q_i(\mu)
\label{eq:h_eff}
\end{equation}
where $G_F$ is the Fermi constant, $C_i(\mu)$ are Wilson 
coefficients taking into account short distance physics and
$Q_i$ are effective local operators. In the Standard Model,
there are 10 operators we need to consider for radiative and
semileptonic decays:
\begin{center}
\begin{tabular}{rclrcl}
${Q_1}$ & ${=} $ & 
${(\bar{s}_i \,c_j)_{V-A}\,(\bar{c}_j \,b_i)_{V-A}}$ &
{ ${Q_2}$} & { ${=} $} & { ${(\bar{s} \,c)_{V-A}
\,(\bar{c} \,b)_{V-A}}$ }\\
${Q_3}$ & ${=} $ & ${(\bar{s} \,b)_{V-A} \sum_q (\bar{q} \,q)_{V-A}}$ &
${Q_4}$ & ${=} $ & ${(\bar{s}_i \,b_j)_{V-A} 
\sum_q (\bar{q}_j \,q_i)_{V-A}}$ \\
${Q_5}$ & ${=} $ & ${(\bar{s} \,b)_{V-A} \sum_q (\bar{q} \,q)_{V+A}}$ &
${Q_6}$ & ${=} $ & ${(\bar{s}_i \,b_j)_{V-A} 
\sum_q (\bar{q}_j \,q_i)_{V+A}}$ \\
{ ${Q_7}$} & { ${=} $ } & { $\frac{{e}}{{8\pi^2}}\, 
{m_b\,\bar{s}_i \,\sigma^{\mu\nu} (1+\dgamma_5)\, b_i\, F_{\mu\nu}}$} &
${Q_8}$ & ${=} $ & $\frac{{g}}{{8\pi^2}}\, {m_b\,\bar{s}_i 
\,\sigma^{\mu\nu} (1+\dgamma_5)\, T^a_{ij}\, b_J\, G^a_{\mu\nu}}$ \\
{ ${Q_9}$} & { ${=} $} 
& { $\frac{{e}}{{8\pi^2}}\, 
{(\bar{s} \,b)_{V-A} \, (\bar{\ell}\,\ell)_V}$} &
{ ${Q_{10}}$} & { ${=} $} & 
{ $\frac{{e}}{{8\pi^2}}\, 
{(\bar{s} \,b)_{V-A} \, (\bar{\ell}\,\ell)_A}$ }
\end{tabular}
\end{center}
The dominant short-distance contributions come from $Q_7$, $Q_9$ and
$Q_{10}$. They arise from penguin and box diagrams illustrated in
Fig.~\ref{fig:PengBox}.  The hadronic matrix elements of these quark bilinears
(Fig.~\ref{fig:2quark}) are straightforward to compute in lattice QCD, at
least for some values of initial and final state momenta.
\begin{figure}[t!]
\null
\vspace{-1ex}
\begin{center}
\includegraphics[clip=true,width=0.25\textwidth]{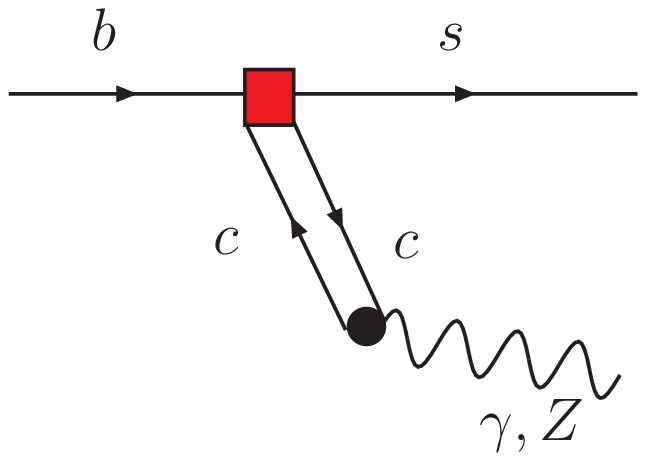}~~~~~
\includegraphics[clip=true,width=0.5\textwidth]{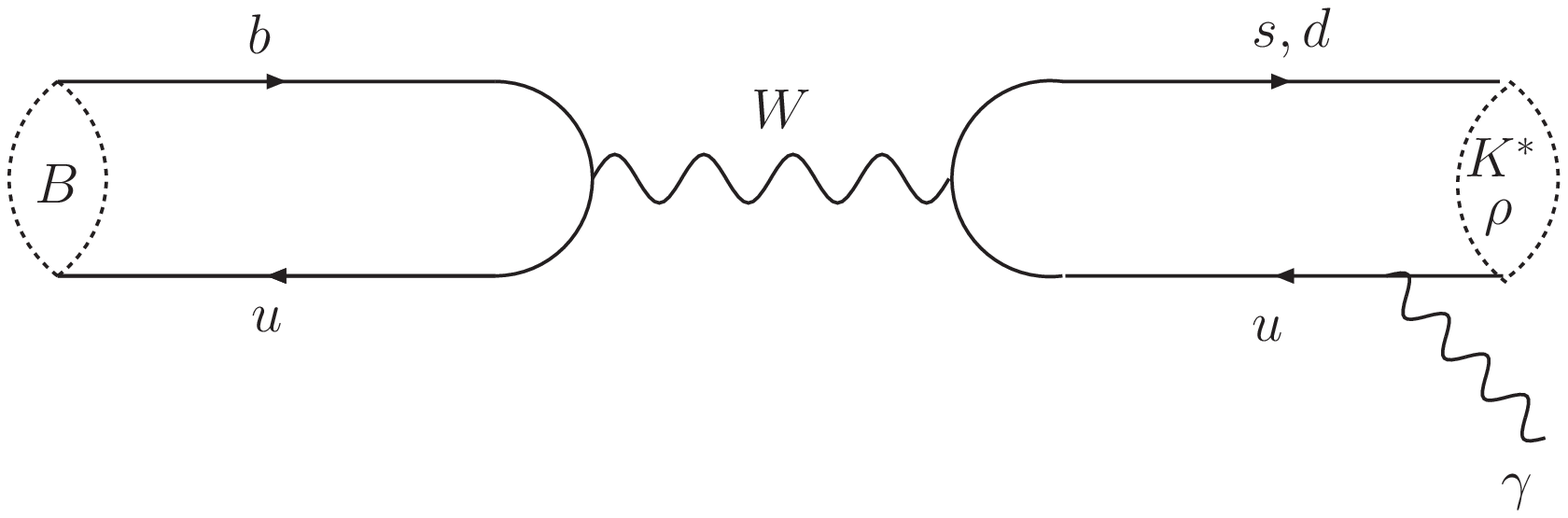}
\end{center}
\vspace{-4ex}
\caption{\label{fig:4quark}Long distance contributions can come
from charmonium resonances via $Q_2$ (left) or weak annihilation (right).
(Weak annihilation is doubly CKM suppressed for $B\to K^*$.)}
\end{figure}
The matrix elements of the quark bilinear currents in $Q_7$, $Q_9$ and
$Q_{10}$ are parametrised by form factors as follows:
\begin{eqnarray}
\langle P(p')|\bar q \dgamma^\mu b |B(p)\rangle &=&
f_+(q^2)\left[p^\mu+p'^\mu-\frac{M_B^2-M_P^2}{q^2}q^\mu\right]
+f_0(q^2)\frac{M_B^2-M_P^2}{q^2}q^\mu,\\
q_\nu\langle P(p')|\bar q \sigma^{\mu\nu} b|B(p) \rangle &=&
\frac{i f_T(q^2)}{M_B+M_P}\left[q^2(p^\mu+p'^\mu)-
(M_B^2-m_P^2)q^\mu\right],\\
\langle V(p',\varepsilon)| \bar q \dgamma^\mu b | B(p) \rangle &=&
 \frac{2iV(q^2)}{M_B+M_V} \,\epsilon^{\mu\nu\rho\sigma}
 \varepsilon^*_\nu \, p'_\rho p_\sigma,\\
\langle V(p',\varepsilon)| \bar q \dgamma^\mu\dgamma_5 b | B(p) 
\rangle &=&
  2M_VA_0(q^2)\,\frac{\varepsilon^*\cdot q}{q^2}\,q^\mu + 
  (M_B+M_V)\,A_1(q^2)\left[\varepsilon^{*\mu}-
  \frac{\varepsilon^*\cdot q}{q^2}\,q^\mu\right]
\nonumber\\[0.0cm]
  &&-\,A_2(q^2)\,\frac{\varepsilon^*\cdot q}{M_B+M_V}
 \left[p^\mu+p'^\mu -\frac{M_B^2-M_V^2}{q^2}\,q^\mu\right],\\
q^\nu\langle V(p',\varepsilon)|\bar q \sigma_{\mu \nu} b |B(p)
\rangle&=&4\:T_1(q^2)\epsilon_{\mu\rho\kappa\sigma}\varepsilon^{*\rho }
p^\kappa p'^\sigma,\\
\nonumber q^\nu\langle V(p',\varepsilon)|\bar q \sigma_{\mu \nu}
\dgamma_5 b |B(p)\rangle&=&2iT_2(q^2)\left[\varepsilon^*_\mu
(M_B^2-M_{V}^2)-(\varepsilon^*\cdot q)(p+p')_\mu \right]\\
&&+2iT_3(q^2)(\varepsilon^*\cdot q)\left[q_\mu
-\frac{q^2}{M_B^2-M_{V}^2}(p+p')_\mu \right],
\end{eqnarray}
where $\varepsilon$ is the polarisation of the vector meson and $q=p-p'$.

The form factors are functions of $q^2$.  We would like to compute
them for the whole range of $q^2$ directly using LQCD;
however it is not possible to do so on current lattice ensembles.
Our calculations are most reliable in the low recoil limit $q^2 \approx
q^2_{\mathrm{max}}$, where both the $B$ meson and the final state meson
are roughly at rest.  The march to high recoil (small $q^2$) is blocked
by three barriers: discretisation errors as spatial momenta in the lattice
frame become comparable to the inverse lattice spacing, errors in
heavy quark effective theory as the QCD dynamics of the decay become
comparable to $m_b$ rather than $\Lambda_{\mathrm{QCD}}$, and growing
statistical errors.

Our strategy for calculating the $q^2$-dependence of the form factors is the
following: (1) Compute with large values of $q^2$ where the form factors above
are the dominant hadronic contributions to $B\to K^{(*)}\ell\ell$ decays.  The
$\ell=\mu$ decays are clean channels at the LHC, and Standard Model
calculations of the differential cross section with $q^2> m_{\bar cc}^2$ will
be tested directly against experimental measurements. (2) Use a combination
of methods, including moving NRQCD, to push LQCD calculations to smaller
values of $q^2$, determining the shape of the form factors over a range
large enough to test various phenomenological ans\"atze for the shapes.
(3) Use these fits to extrapolate to the low $q^2$ region, down to
$q^2=0$.

We have discussed above the contributions from $Q_7$, $Q_9$, and
$Q_{10}$.  Compared to these, contributions from $Q_1$ and $Q_{3-6}$ are
loop-suppressed.  The contribution from $Q_8$ is also suppressed by
an approximate factor $(\Lambda_{\mathrm{QCD}}/m_B)(C_8/C_7)\approx 0.05$.
There are, however, complications which have to be understood due to $Q_2$.
The contribution from $Q_2$ dominates when the $\bar cc$ pair is resonant
(Fig.~\ref{fig:4quark}, left). This is non-local and we can not calculate it
on the lattice.  Therefore, we will not be able to extract the differential
branching fraction in ranges of $q^2$ which are close to a charmonium
resonance.  Nevertheless, the form factors we do calculate will be unaffected
by these resonances and extrapolations in $q^2$ will be smooth.

Weak annihilation contributions, where the valence quarks of the $B$
meson annihilate into a $W$ which decays into a $u$ quark and 
a $d$ or $s$ quark (Fig.~\ref{fig:4quark}, right), are long-distance
contributions to radiative and semileptonic $B \to K^{(*)}$ decays,
since the photon or $Z$ is emitted separately from the flavour-changing
interaction.  This process is highly suppressed for $b\rightarrow s$ 
decay since $V_{ub}V_{us}^*\ll V_{tb}V_{ts}^*$, in contrast to $b\to d$
decay where there is no CKM suppression.

Another complication facing LQCD calculations with vector mesons in the final
state is their instability to strong decays.
With the presently used value for the light quark mass, the $\rho$
and $K^*$ are stable, but eventually we need to extrapolate to the
physical light quark mass. The effects at the decay threshold can be perhaps
be studied in separate work by looking at the vector meson decay constant as
the quark mass is reduced. We can also try to check the validity of the
extrapolation by comparing $|V_{ub}|$ obtained from $B\rightarrow \rho \ell
\nu$ with that from $B\rightarrow \pi \ell \nu$.

\section{Moving NRQCD and matching of heavy-light operators}
\label{sec:current_matching}

To reduce discretisation errors for the light meson in the final state
at high recoil, we work in a reference frame where the $B$ meson is
not at rest, so that for a given value of $q^2$ the
momentum of the light meson is reduced. Moving nonrelativistic QCD (mNRQCD)
allows us to treat the momentum of the heavy quark arising from the
frame choice exactly.

The full $\mathcal{O}(\Lambda_{QCD}^2/m_b^2)$ mNRQCD action used in
our calculation was derived in Ref.~\cite{Horgan:2009ti}. Schematically,
the 4-momentum of the $b$ quark is parametrised as $p=m_b u+k$,
where $u=\gamma(1,\vec v)$ and $\gamma=1/\sqrt{1-v^2}$. Then a non-relativistic
expansion in the 3-momentum $\vec k$ is performed.

At tree-level, the QCD heavy quark field $\Psi(x)$ is related to the
mNRQCD two-component quark- and antiquark fields $\psi_v(x)$, $\xi_v(x)$
by the transformation
\begin{equation}
\Psi(x)=S(\Lambda)T_{FWT}e^{-im u\cdot x\dgamma^0}
A_{D_t}\frac{1}{\sqrt{\gamma}}\Psi_v(x)\hspace{4ex}{\rm with}
\hspace{4ex}\Psi_v(x)=\left(\begin{array}{c}\psi_v(x)\\
\xi_v(x)\end{array}\right),
\label{eq:mNRQCD_field_redef}
\end{equation}
where $e^{-im u\cdot x\dgamma^0}$ removes the additive heavy quark mass 
term, $T_{FWT}$ is the Foldy-Wouthuysen-Tani transformation,
$S(\Lambda)$ is the spinorial representation of the boost and $A_{D_t}$ removes
time derivatives in the Hamiltonian.

Eq.~(\ref{eq:mNRQCD_field_redef}) can be used as a starting point for the construction of the
lattice operators corresponding to the heavy-light currents in the operators $Q_7$,
$Q_9$ and $Q_{10}$. For a current of the form $J=\bar{q} \Gamma b$ with some Dirac
matrix $\Gamma$, one obtains, using the equations of motion to eliminate
time derivatives,
\begin{equation}
J=\frac{1}{\sqrt{\gamma}}\:\bar{q}\:\:\Gamma\:\:S(\Lambda)\:\Psi_v^{(+)}
+ \frac{1}{2m_b\sqrt{\gamma}}\:\bar{q}\:\:\Gamma\:\:(-i\dgamma^0 \vec{v}
+ i\vec{\dgamma}  + i \vec{v}/\gamma  )\cdot\vec{D} \:\:S(\Lambda)
\:\Psi_v^{(+)} + \mathcal{O}(1/m_b^2) \label{eq:tree_level_current}
\end{equation}
where $\Psi_v^{(+)}$ has vanishing lower components. On the lattice, the continuum covariant
derivative $\vec{D}$ has to be replaced by a discrete version $\vec{\Delta}$.
For the light/strange quark $q$, we use an improved staggered action; presently ASQTAD. This means
that the field $q$ has to be expressed in terms of the staggered field $\chi_q$.

\begin{figure}[t!]
\null
\vspace{-4ex}
\begin{minipage}[t]{.48\linewidth}
\centerline{\includegraphics[width=0.6\linewidth,angle=270]{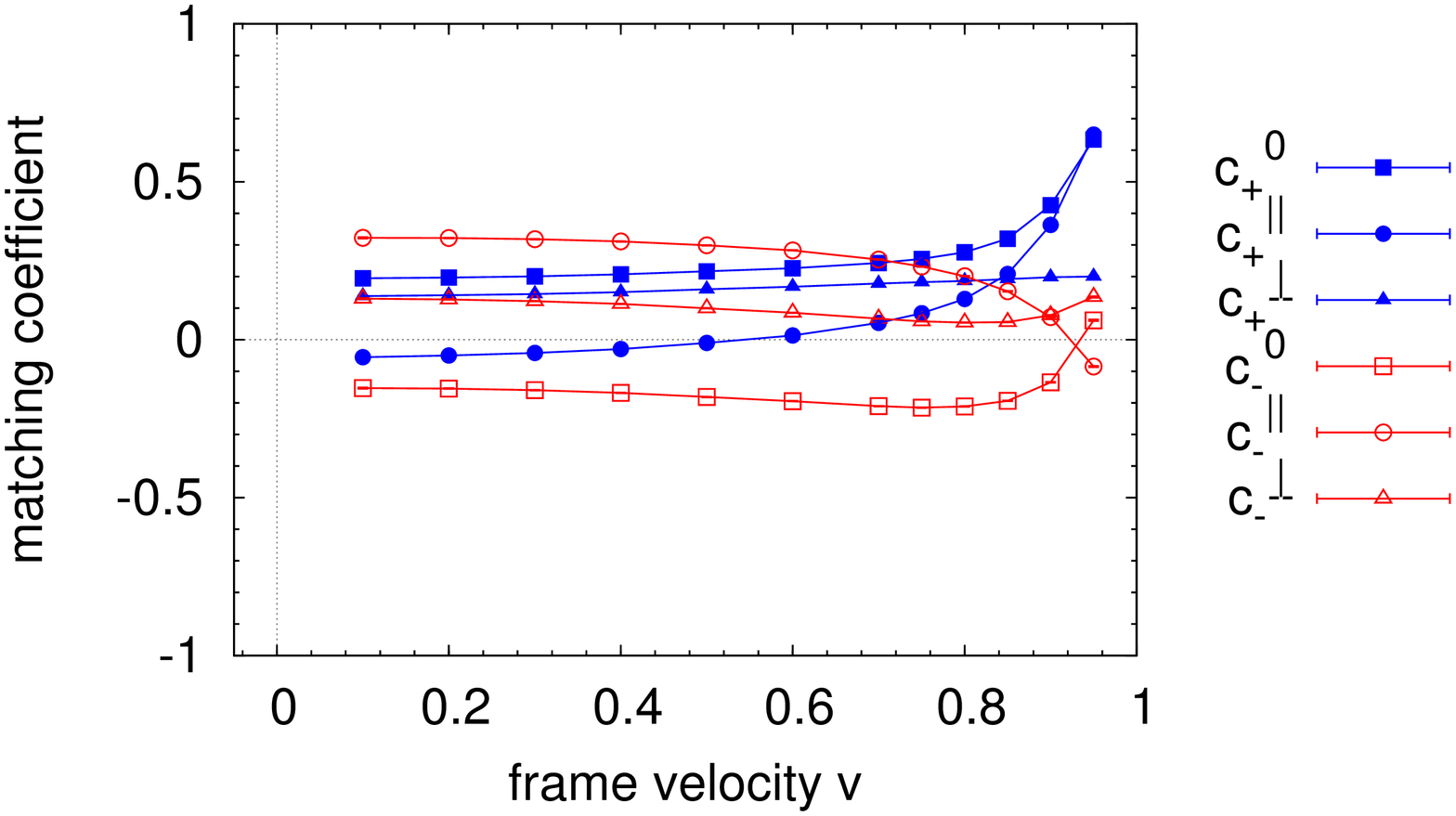}}
\caption{Matching coefficients for the vector current.}
\label{fig:plot_match_full_vector}
\end{minipage}
\hfill
\begin{minipage}[t]{.48\linewidth}
\centerline{\includegraphics[width=0.6\linewidth,angle=270]{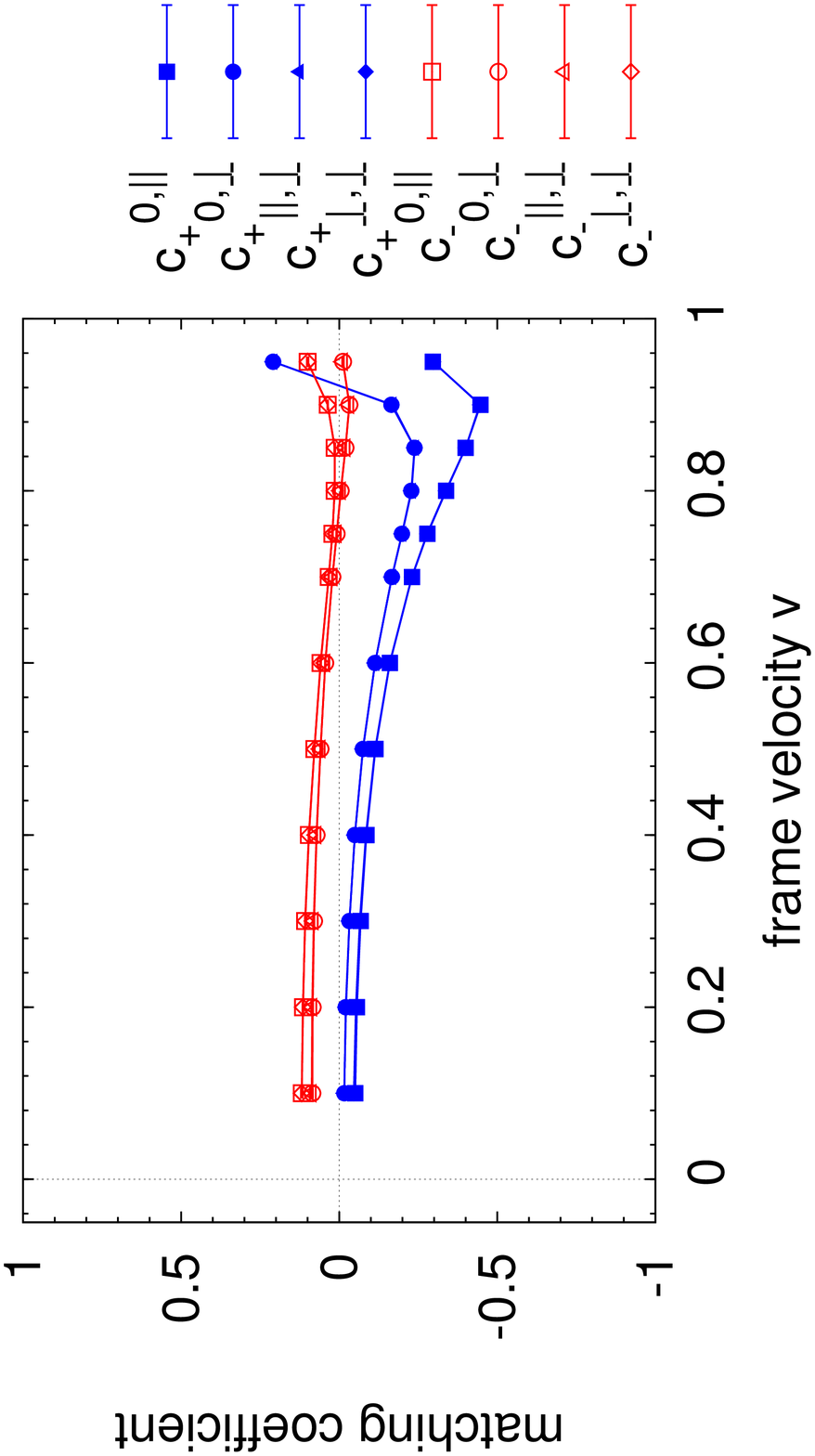}}
\caption{Matching coefficients for the tensor current. The renormalisation scale is $\mu=m_b$.}
\label{fig:plot_match_full_tensor}
\end{minipage}
\end{figure}

Eq.~(\ref{eq:tree_level_current}) does not yet include any radiative corrections.
Since the second term in (\ref{eq:tree_level_current}) is suppressed by $\Lambda_{QCD}/m_b$
relative to the first term, we currently restrict our calculation of radiative corrections
to the first term. To obtain these, we note that the spinorial boost is given explicitly by
\begin{equation}
 S(\Lambda)=S_+(\Lambda)=\frac{1}{\sqrt{2(1+\gamma)}}\left[(1+\gamma)\hat{1}
- \gamma \:\vec{v}\cdot\vec{\dgamma}\:\dgamma^0 \right]. \label{eq:S_lambda}
\end{equation}
As can be seen, (\ref{eq:S_lambda}) contains a sum of two different Dirac structures; these
will mix under renormalisation. Thus, at order $\mathcal{O}(\alpha_s)$, we also need to
consider the combination with the opposite sign:
\begin{equation*}
 \phantom{S(\Lambda)=}S_-(\Lambda)\equiv\frac{1}{\sqrt{2(1+\gamma)}}\left[(1+\gamma)\hat{1}
+ \gamma \:\vec{v}\cdot\vec{\dgamma}\:\dgamma^0 \right].
\end{equation*}
The lattice current through order $\mathcal{O}(\alpha_s)$ then reads
\begin{equation}
J^{(0)\:\rm lat}=(1+\alpha_s\:c_+)\:J^{(0)}_{+} + \alpha_s \:c_-\:\:J^{(0)}_{-} \label{eq:lattice_current}
\end{equation}
with
\begin{equation*}
 J^{(0)}_{+} = \frac{1}{\sqrt{\gamma}}\:\bar{q}\:\:\Gamma\:\:S_+(\Lambda) \:\Psi_v^{(+)}, \hspace{4ex}
 J^{(0)}_{-} = \frac{1}{\sqrt{\gamma}}\:\bar{q}\:\:\Gamma\:\:S_-(\Lambda) \:\Psi_v^{(+)}.
\end{equation*}
The matching coefficients $c_{\pm}$ in (\ref{eq:lattice_current}) are obtained by requiring
that the matrix elements in the lattice theory and in the continuum ($\overline{\mbox{MS}}$ scheme)
are equal to order $\alpha_s$.

Results for the matching coefficients for the vector ($\Gamma=\dgamma^\mu$) and tensor
($\Gamma=\sigma^{\mu\nu}$) currents are shown in Figs.~\ref{fig:plot_match_full_vector}
and \ref{fig:plot_match_full_tensor}. These were obtained using automated tadpole-improved
one-loop lattice perturbation theory ~\cite{Muller:2009af}. Here, the bare $b$ quark mass
was set to $am_b=2.8$ and the boost-velocity is pointing in 1-direction. In the figures,
the different Lorentz indices of $\Gamma$ are indicated as $0$ (temporal),
$\|$ (parallel to $\vec{v}$) and $\perp$ (perpendicular to $\vec{v}$).

The tree-level $\mathcal{O}(\Lambda_{QCD}/m_b)$ correction is
\begin{equation}
 J^{(1)}_{+}= \frac{1}{m_b}\frac{1}{\sqrt{\gamma}}\:\bar{q}\:\:\Gamma\:\frac{ (-i\dgamma^0 \vec{v}
+ i\vec{\dgamma}  + i \vec{v}/\gamma  )\cdot\vec{\Delta} }{2}\:\: S_{+}(\Lambda)
\:\Psi_v^{(+)}.
\label{eq:tree_level_1m_current}
\end{equation}
Because of the mixing-down that occurs, one should work with subtracted $1/m$ currents
in the calculation of the form factors. We have already computed the non-perturbative matrix
elements of (\ref{eq:tree_level_1m_current}), so that we can include them
in the form factor results once we have obtained the necessary mixing coefficients
from perturbation theory.

\section{Two-point and three-point functions with stochastic sources}
\label{sec:stochstic_sources}

In order to extract the form factors, we need to compute two-point functions for the
light meson and the $B$ meson, as well as three-point functions with the currents
$J^{\rm lat}$ (see section \ref{sec:current_matching}) inserted. The basic method for the
extraction of the matrix elements from fits to correlators was outlined in~\cite{Meinel:2008th}.

Here, we investigate the use of stochastic (random wall) sources, which allow us
to obtain approximations to the all-to-all correlators and therefore possibly to reduce
statistical errors. For the $B\rightarrow\pi \ell\nu$ semileptonic decay
in standard NRQCD (i.e with the $B$ meson at rest), random wall sources have
been tested in Ref.~\cite{Davies:2007vb}. We now include vector meson final
states and moving NRQCD in our study, and also work at smaller light-quark masses.

For the light valance quarks, we use one-spinor-component staggered fermions.
We work with four-spinor-component naive quarks \cite{Wingate:2002fh} in constructing
the interpolating fields for heavy-light and light-light mesons. The naive quark propagator
$G_q(y,x)$ is related to the one-spinor-component staggered quark
propagator $G_{\chi_q}(y,x)$ by
\begin{equation}
G_q(y,x)=G_{\chi_q}(y,x)\Omega(y)\Omega^\dagger(x),
\end{equation}
where $\Omega(x)=\prod_{\mu=0}^{3}(\dgamma_\mu)^{x_\mu}$.

The numerical calculations described in the following sections were performed
using 400 ``coarse'' MILC gauge configurations ~\cite{Aubin:2004wf}
that have $V=20^3\times64$ and $a^{-1}\approx 1.6$ GeV. These configurations
include $2+1$ flavours of ASQTAD sea quarks with masses $am_u=am_d=0.007$ and $am_s=0.05$.
The ASQTAD action was also used for the light valence quarks, with masses $am_u=am_d=0.007$ and
$am_s=0.04$. The heavy quark mass and stability parameter were set to $am_b=2.8$ and $n=2$.

\subsection{Light meson two-point functions}

\label{sec:ll2pt}

For flavour non-singlet light pseudoscalar mesons with 3-momentum $\vec p'$, 
the exact all-to-all two-point function (with naive quarks) is given by
\begin{eqnarray}
C_{55}(t=y_0-x_0,\vec p')&\equiv&\frac{1}{L^3}\sum_{\vec y,\vec x}
\langle \Phi_5(y)\Phi_5^\dagger(x)\rangle e^{-i\vec p'\cdot(\vec y-\vec x)}
\nonumber\\
&=&\frac{1}{L^3}\sum_{\vec y, \vec x}\Tr[\dgamma_5 G_q(y,x)\dgamma_5
\dgamma_5 G_{q'}^\dagger(y,x)\dgamma_5]e^{-i\vec p'\cdot(\vec y-\vec x)}
\nonumber\\
&=&\frac{4}{L^3}\sum_{\vec y, \vec x}\Tr[G_{\chi_q}(y,x)G_{\chi_{q'}}^\dagger(y,x)]
e^{-i\vec p'\cdot(\vec y-\vec x)}.
\label{eq:ps_2pt_all_to_all}
\end{eqnarray}
Note that due to the taste doubling, the amplitude obtained from this correlator
is too large by a factor of 16 (in the continuum limit), which needs to be divided out.

Compared to a single-point-source correlator, where one does not sum over $\vec x$,
we would expect a reduction in the statistical errors by a factor 
proportional to
$\sqrt{m_\pi^3 L^3}$. However, it is forbiddingly expensive to compute (\ref{eq:ps_2pt_all_to_all})
directly. Instead, one can approximate (\ref{eq:ps_2pt_all_to_all}) using random wall sources, as
explained in the following.

We define
\vspace{-2ex}
\begin{equation*}
\widetilde{G}_{\chi_q}^p(y,x_0,\vec{p'})=\sum_{\vec{x}}\:G_{\chi_q}(y,x)\:\xi^p(\vec{x})
\:e^{i\vec p' \cdot\vec{x}}
\end{equation*}
where $\xi^p(\vec{x})$ is a vector in colour space (colour index not shown explicitly),
with every colour component at every spatial site an independent $Z_2\times Z_2$ random number
(note that in \cite{Davies:2007vb} a different type of noise was used).
The index $p$ labels different random samples on a given configuration; additionally it
is understood that new random numbers are used on every gauge configuration. The noise fields satisfy
\vspace{-1.5ex}
\begin{equation}
\frac{1}{n_Z}\sum_{p=1}^{n_Z}\xi_c^{p*}(\vec x)\xi_d^{p}(\vec z)\approx\delta_{cd}
\delta_{\vec x\:\:\vec z}
\end{equation}
where $n_Z$ is the number of random samples (this relation becomes exact in the limit
$n_Z \rightarrow \infty$). Therefore, an approximation to the all-to-all pseudoscalar
two point function can be obtained as follows:
\vspace{-1.5ex}
\begin{eqnarray}
C_{55,{\rm RW}}(t,\vec p')&=&\frac{4}{L^3}\frac{1}{n_Z}\sum_{p=1}^{n_Z}\sum_{\vec y}
\widetilde{G}_{\chi_{q'}}^p(y,x_0,\vec{0})^*\cdot
\widetilde{G}_{\chi_q}^p(y,x_0,\vec{p'}) e^{-i\vec p'\cdot \vec y} \label{eq:ps_2pt_RW} \\
\nonumber &\approx&\frac{4}{L^3}\sum_{\vec y,\vec z,\vec x}
G_{\chi_{q'}}^*(y,z)_{ab}\:
G_{\chi_q}(y,x)_{ac}\:\delta_{bc}\:\delta_{\vec x\:\:\vec z}\:e^{i\vec p' \cdot\vec{x}}
e^{-i\vec p'\cdot \vec y}=C_{55}(t,\vec p').
\end{eqnarray}
In practice, a small number $n_Z$ is sufficient; even $n_Z=1$ works to improve
the signal for pseudoscalars compared to point sources. However, note that for each
non-zero value of the momentum $\vec p'$, new inversions are required.

For a flavour non-singlet vector meson with interpolating field
$\Phi_j=\bar q'\dgamma_j q$, the exact all-to-all two point function is given by
\vspace{-1.5ex}
\begin{eqnarray}
C_{jj}(t,\vec p')&\equiv&\frac{1}{L^3}\sum_{\vec y,\vec x}
\langle \Phi_j(y)\Phi_j^\dagger(x)\rangle e^{-i\vec p'\cdot(\vec y-\vec x)}
\nonumber\\
&=&\frac{1}{L^3}\sum_{\vec y, \vec x}\Tr[\dgamma_j G_q(y,x)\dgamma_j
\dgamma_5 G_{q'}^\dagger(y,x)\dgamma_5]e^{-i\vec p'\cdot(\vec y-\vec x)}
\nonumber\\
&=&\frac{4}{L^3}\sum_{\vec y, \vec x}\Tr[G_{\chi_q}(y,x)G_{\chi_{q'}}^\dagger(y,x)]
(-1)^{x_j+y_j} e^{-i\vec p'\cdot(\vec y-\vec x)},
\end{eqnarray}
where the phase factor of $(-1)^{x_j+y_j}$ comes from the relation 
$\Omega^\dagger(x)\dgamma_5\dgamma_j\Omega(x)=
(-1)^{x_j}\dgamma_5\dgamma_j$.

To obtain the random wall correlator in this case, a factor of $(-1)^{x_j}$ is added to the 
stochastic source for the zero-momentum quark propagator; we define
\begin{equation}
\widetilde{G}_{\chi_{q'}}^p(y,x_0,j)=\sum_{\vec{x}}\:G_{\chi_{q'}}(y,x)\:\xi^p(\vec{x})\:(-1)^{x_j}.
\end{equation}
This means that additional inversions are required for the different polarisations $j=1,2,3$.
The random wall correlator is then
\begin{equation}
C_{jj,{\rm RW}}(t,\vec p')= \frac{4}{L^3}\frac{1}{n_Z}\sum_{p=1}^{n_Z}\sum_{\vec y}
\widetilde{G}_{\chi_{q'}}^p(y,x_0,j)^* \cdot
\widetilde{G}_{\chi_q}^p(y,x_0,\vec{p'}) (-1)^{y_j} e^{-i\vec p'\cdot\vec y}.
\label{eq:vector_2pt_RW}
\end{equation}
A modification of the simple random wall sources is colour dilution, where a Kronecker delta
in colour space is introduced in the stochastic source, i.e. $\xi_a^p(\vec z)$ is
changed to $\delta_{a,a_0}\xi^p(\vec z)$ for each source colour $a_0$. Here,
$\xi^p(\vec z)$ does not have a colour index.
Then for each colour $a_0$ one needs to perform a separate inversion and in 
Eqs.~(\ref{eq:ps_2pt_RW},\:\:\ref{eq:vector_2pt_RW}) a sum over source colour has to be added.

\begin{figure}[t!]
\begin{center}
\includegraphics[clip=true,width=0.45\textwidth,height=2in]{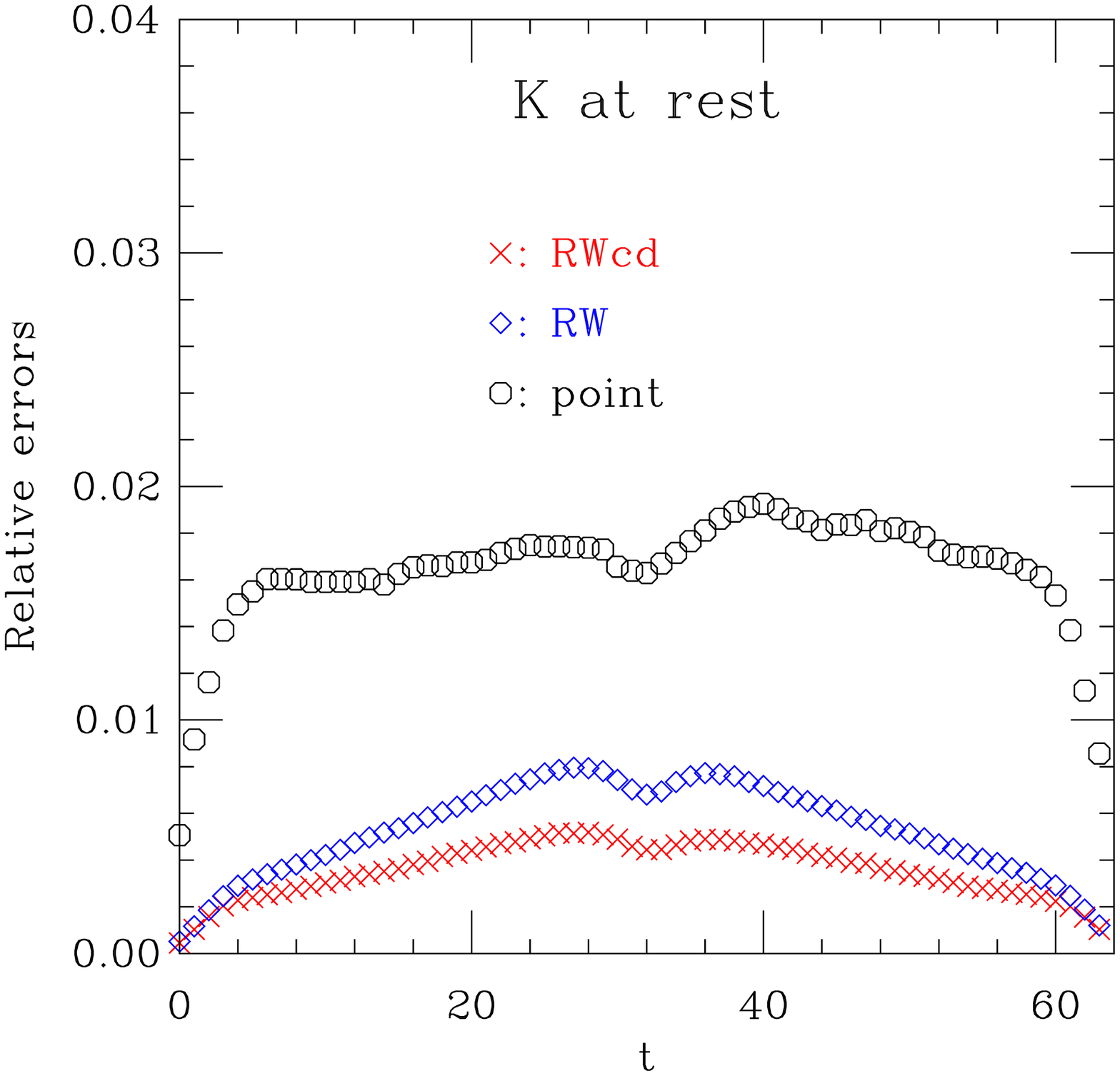}
\hfill
\includegraphics[clip=true,width=0.45\textwidth,height=2in]{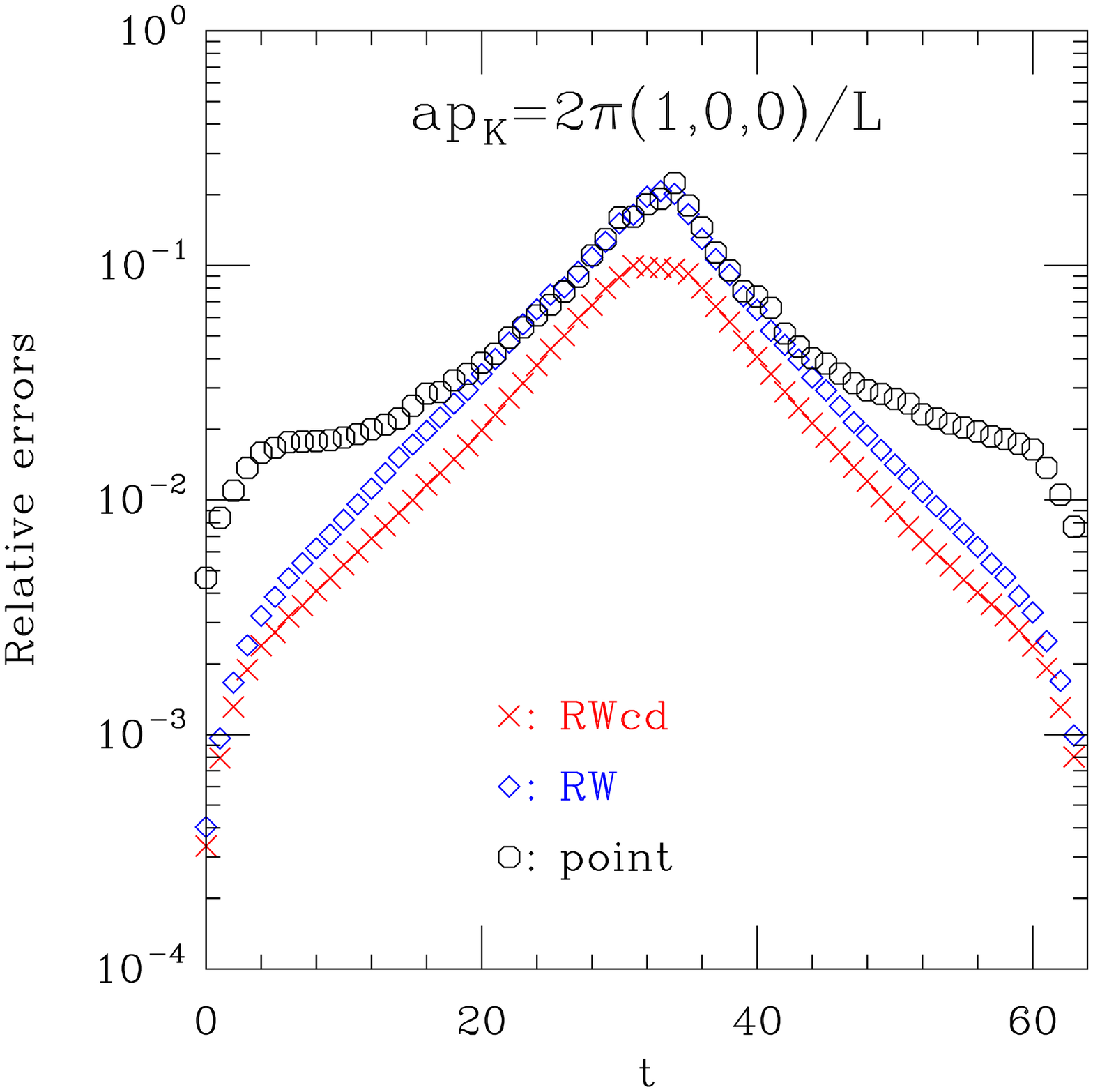}
\end{center}
\caption{\label{fig:errK}Comparison of relative errors of $K$ meson
two point functions from random wall sources (RW) and a point 
source. "RWcd" means random wall sources with colour dilution.
See the main text for the numbers of inversions used for the different source types.}
\end{figure}

Fig.~\ref{fig:errK} shows the comparison of the relative
errors of $K$ meson two-point functions from random wall sources and
a point source at zero momentum and at $ap=2\pi(1,0,0)/L$. "RWcd" refers to
the random wall source with colour dilution and "RW" the simple random
wall source. The statistical errors are significantly reduced by using
random wall sources. However the improvement decreases as the momentum 
of the meson increases. We put four sources on each configuration
for each type of sources. Therefore the number of inversions for the RWcd
and point source is the same ($1600\times3$) while three times fewer inversions
are used for the standard RW source.
We use the Bayesian fitting method described in Ref.~\cite{Lepage:2001ym} to fit the kaon
two-point functions with a function of the form
\begin{equation}
C_{55}(\:t,\vec p')=\sum_{n=0}^{N-1}(A_n^{55})^2 \left[ e^{-E_n t}+e^{-E_n(L_t-t)}\right]
+ (-1)^{t+1}\sum_{n=0}^{\widetilde{N}-1}(\widetilde{A}_n^{55})^2 \left[ e^{-\widetilde{E}_n t}+e^{-\widetilde{E}_n(L_t-t)}\right],
\label{eq:fit_func_c55}
\end{equation}
where oscillating terms are included for the parity partners. To ensure the correct ordering of the states, we
actually use the logarithms of the energy differences, $\ln (E_n-E_{n-1})$, $\ln (\widetilde{E}_n-\widetilde{E}_{n-1})$ as the fit parameters. Also, the amplitudes of the excited
states ($n>0$) are written as $A_n=B_n\:A_0$,\hspace{1ex} $\widetilde{A}_n=\widetilde{B}_n\:\widetilde{A}_0$, and the
relative amplitudes $B_n$, $\widetilde{B}_n$ are used as the fit parameters. The numbers of exponentials
$N$ and $\widetilde{N}$ are increased until the results are stable. For the kaon at zero momentum, one
can set $\widetilde{N}=0$.

\begin{figure}[t!]
\begin{center}
\includegraphics[clip=true,width=0.49\textwidth]{K_mom_0_0_0_bay5_A2_t2}
\hfill
\includegraphics[clip=true,width=0.49\textwidth,height=2in]{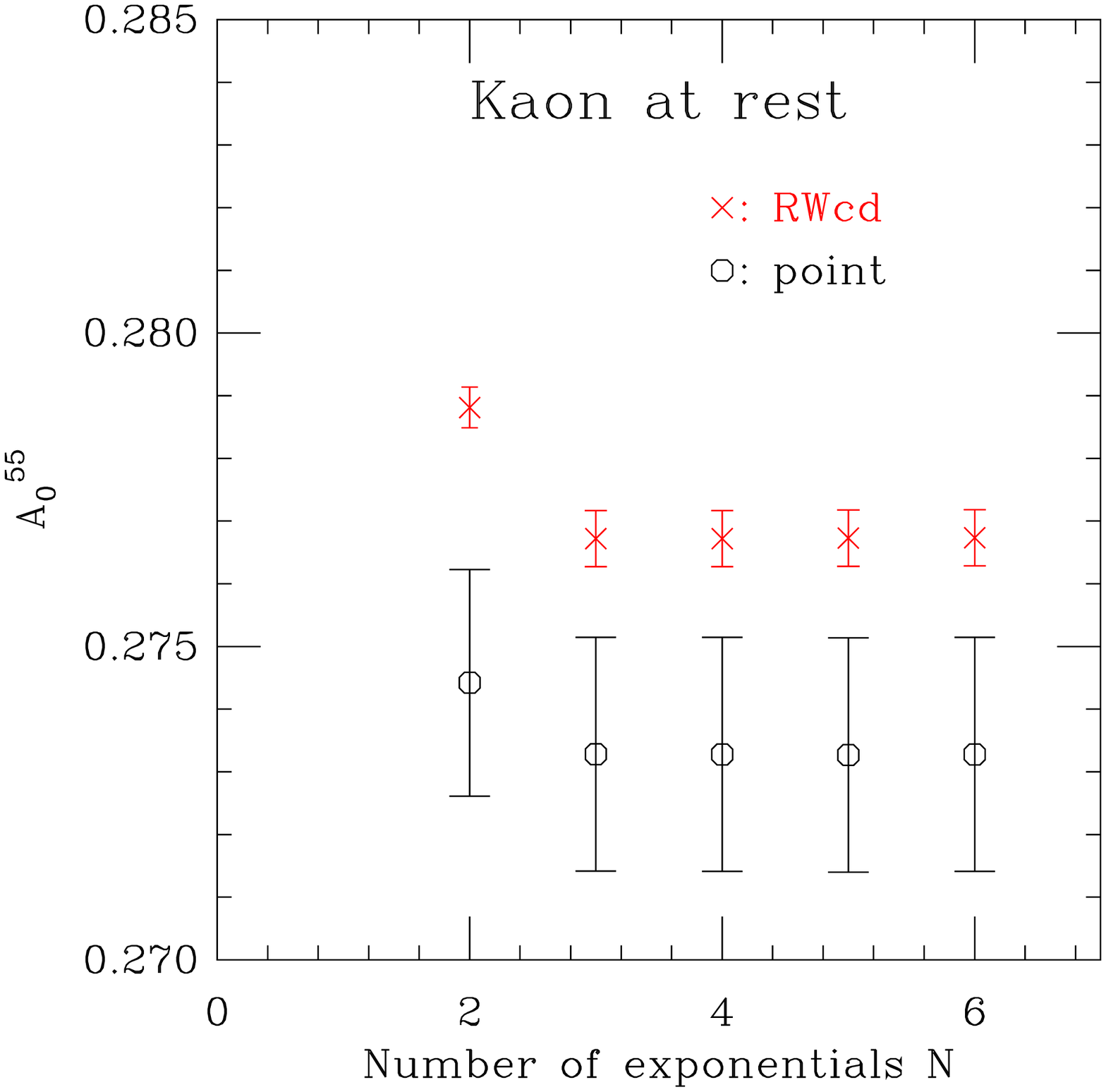}
\end{center}
\caption{\label{fig:bfitK000}Left: A constrained five 
exponential fit
to the $K$ meson correlator from the random wall source with colour
dilution (fit range: $t=2...62$). Right: Fit values for the ground
state amplitude against the number of exponentials $N$ in the 
fits, for the RWcd and point sources.}
\end{figure}

An example fit to the $K$ meson two-point function at zero momentum
with the RWcd source is shown in Fig.~\ref{fig:bfitK000} (left-hand side).
On the right-hand side of Fig.~\ref{fig:bfitK000}, we compare the results for
the ground state amplitude from the RWcd and point sources.
We observe a factor of 4.2 improvement in the ground state amplitude and 1.4 in the energy.
For the kaon with momentum $ap=2\pi(1,0,0)/L$, these improvement factors
are 3.6 and 2.6, respectively.
There is a small (about 1 percent or 1.8$\sigma$) deviation in the amplitudes between the RWcd
and point source correlators. Note that we inadvertently used different temporal boundary conditions
for the two types of sources, but we would expect this to have an effect much smaller than 1 percent.

\begin{figure}[t!]
\begin{center}
\includegraphics[width=0.45\textwidth,height=2in]{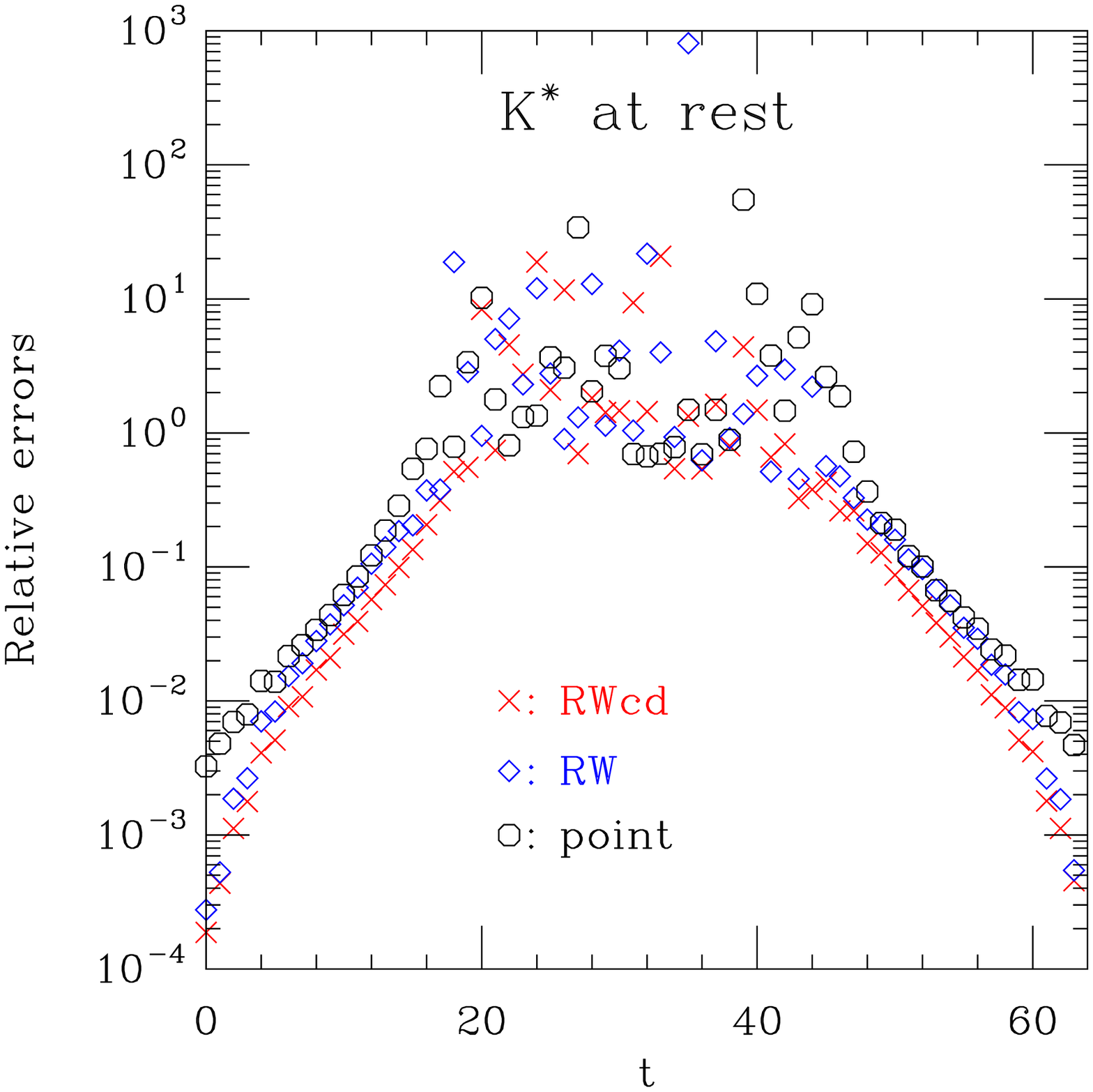}
\hfill
\includegraphics[clip=true,width=0.45\textwidth,height=2in]{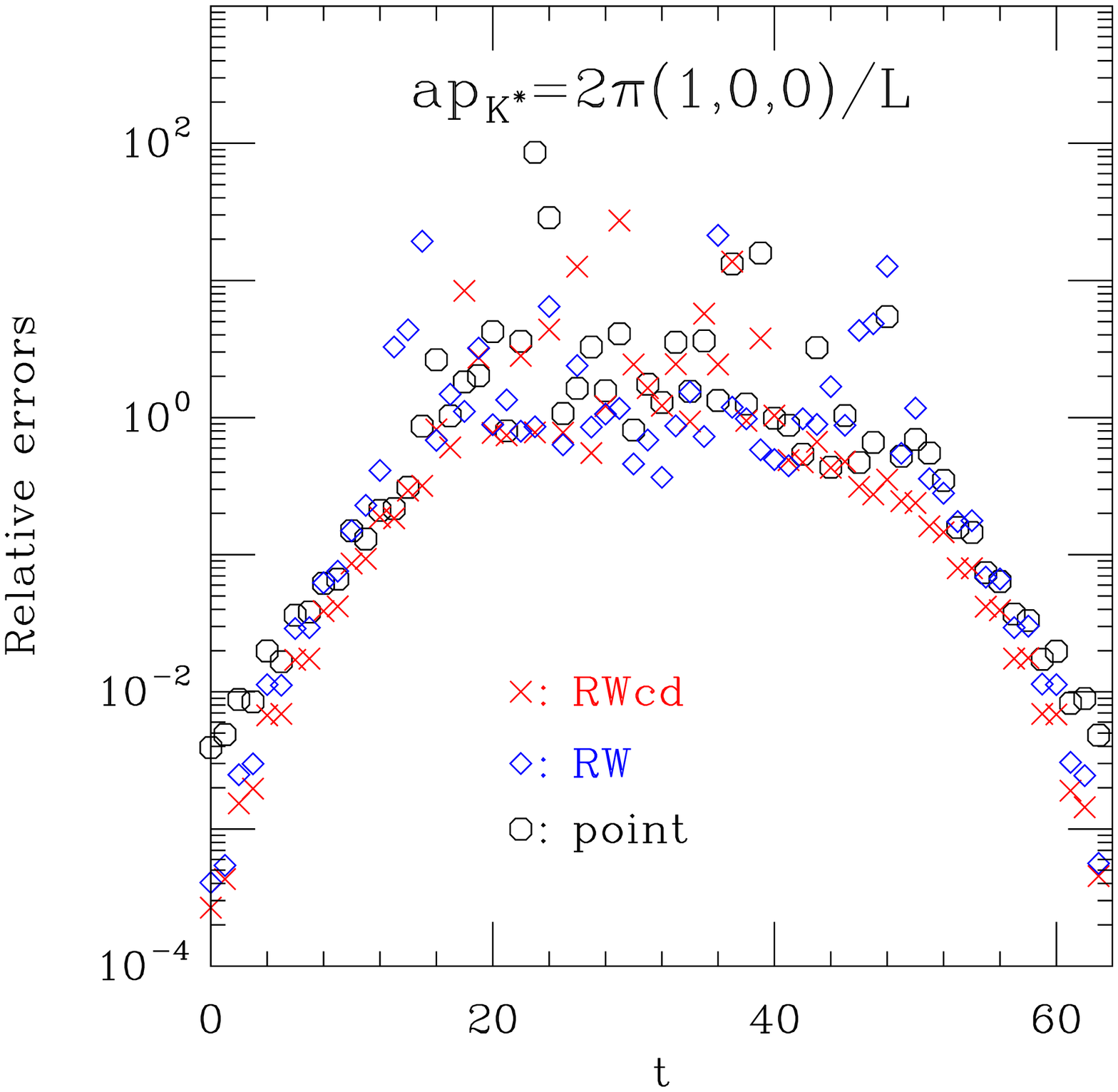}
\end{center}
\caption{\label{fig:errKstar}Comparison of relative errors of $K^*$
meson two point functions from random wall and point sources. See the main text for the
numbers of inversions used for the different source types.}
\end{figure}

For the vector meson correlator with zero momentum, we average
all three polarisations.
With nonzero momentum in the $\hat x$-direction, only the two
transverse polarisations are averaged. 
In Fig.~\ref{fig:errKstar} we compare the relative
errors of $K^*$ two-point functions from random wall sources and a point 
source at zero momentum and at $ap=2\pi(1,0,0)/L$.
\begin{figure}[t!]
\begin{center}
\includegraphics[clip=true,width=0.49\textwidth]{Kstar01_xyz_re_mom_0_0_0_bay55_A2_t1}
\hfill
\includegraphics[clip=true,width=0.49\textwidth,height=2in]{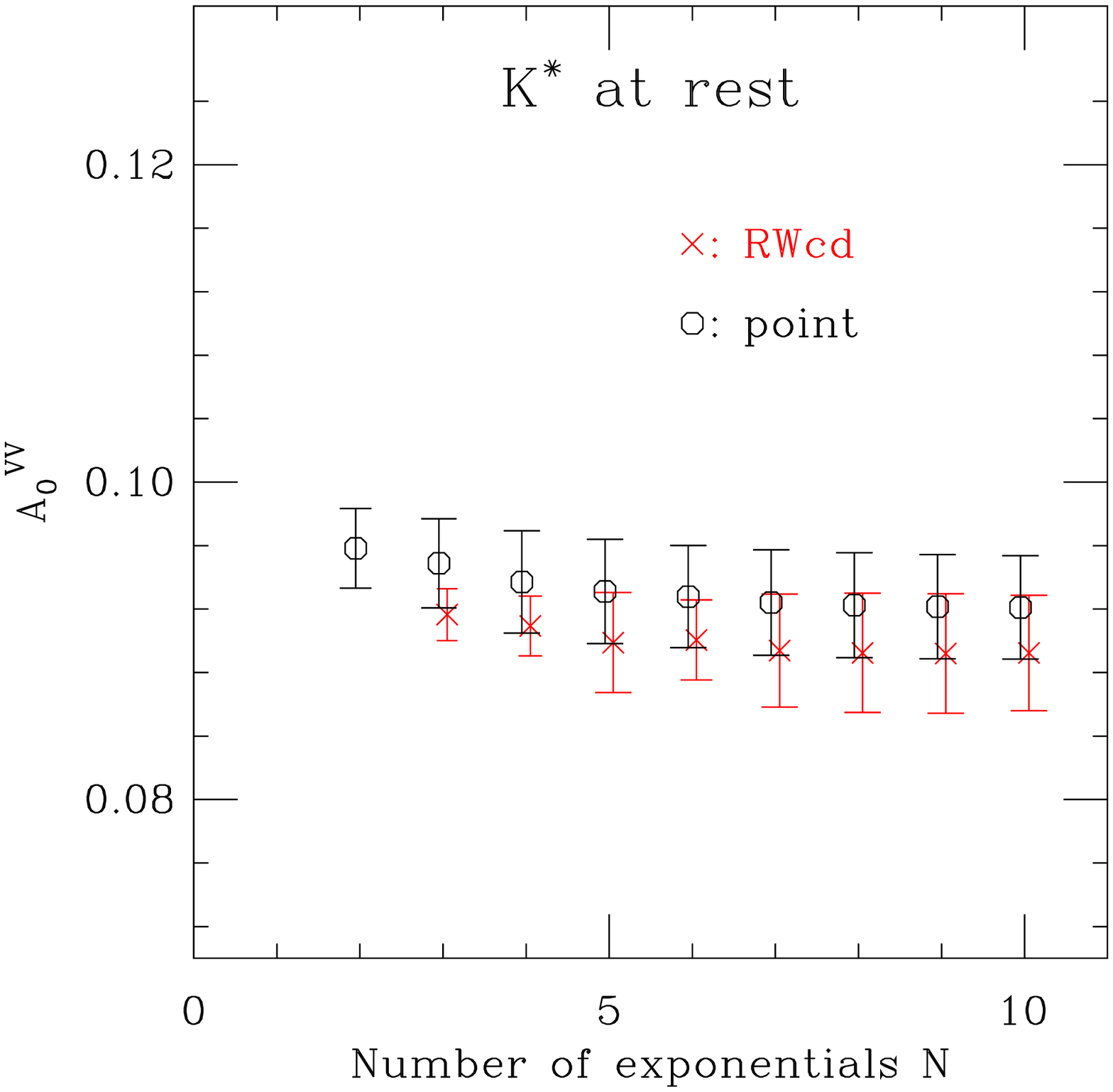}
\end{center}
\caption{\label{fig:bfitKstar000}Left: A constrained five 
exponential fit
to the $K^*$ meson correlator from the random wall source with colour
dilution (fit range: $t=1...63$). Right: Fit values for the ground
state amplitude against the number of exponentials $N=\widetilde{N}$
in the fits, for the RWcd and point sources.}
\end{figure}
The number of inversions for the RWcd, RW and point sources are
$1600\times3\times4$, $1600\times4$ and $1600\times3$, respectively, for
zero momentum. For a nonzero momentum, the numbers are 
$1600\times3\times3$, $1600\times3$ and $1600\times3$, respectively.

As can be seen in Fig.~\ref{fig:errKstar}, for the vector meson two-point functions
a reduction in the statistical errors is only seen at small $t$.
An example fit for the $K^*$ at rest with the RWcd source is shown in Fig.~\ref{fig:bfitKstar000} (left-hand side).
On the right-hand side of Fig.~\ref{fig:bfitKstar000}, we compare the results for
the ground state amplitude from the RWcd and point sources. Unfortunately,
no improvement is observed, even though the computational cost for the RWcd source was higher.

\subsection{Heavy-light two-point functions}
\label{sec:hl2pt}

For $B/B_s$ mesons, the exact all-to-all correlator with moving NRQCD and staggered
actions is (for $t=y_0-x_0>0$)
\begin{equation}
 C_B(t,\vec{k})=\frac{1}{\gamma}\frac{1}{L^3}\sum_{\vec{y},\vec{x}}
\mathrm{Tr}\left[\:G_{\chi_{q'}}^\dag(y,x)\:\Omega^\dag(y)\:S(\Lambda)
\left(\begin{array}{cc} G_{\psi_v}(y,x) & 0 \\ 0 & 0\end{array}\right)
\overline{S}(\Lambda)\:\Omega(x)\:\right]e^{-i\vec{k}\cdot(\vec{y}-\vec{x})}.
\end{equation}
Note that due to the use of mNRQCD for the $b$ quark, the physical momentum
$\vec{p}$ is related to the lattice momentum $\vec{k}$ by
\begin{equation}
\vec{p}=\vec{k}+Z_p\:\gamma\: m_b \vec{v},\\
\end{equation}
where $Z_p\approx 1$ is the renormalisation of the external momentum.
Similarly, the physical energy $p_0=E_B$ of the $B$ meson is related
to the energy $E_v(\vec{k})$ obtained from the fit by
\begin{equation}
E_B(\vec{p})=E_v(\vec{k})+C_v
\end{equation}
where $C_v$ is the velocity-dependent energy shift. Both $C_v$ and $Z_p$ have been
calculated both non-perturbatively and perturbatively in Ref.~\cite{Horgan:2009ti}.

In order to obtain the random wall correlator, we define
\begin{equation}
 \widetilde{G}_H^p(y,x_0,\vec{k})=\sum_{\vec{x}}\:\left(\begin{array}{cc}
G_{\psi_v}(y,x) & 0 \\ 0 & 0\end{array}\right)\overline{S}(\Lambda)
\:e^{i\vec{k}\cdot\vec{x}}\:\Omega(x)\:\xi^p(\vec{x}). \label{eq:B_src}
\end{equation}
We then combine this with the zero-momentum random wall light quark propagator from
the same $\xi^p$ to obtain the $B$ meson random-wall correlator:
\begin{equation}
 C_{B,\rm RW}(t,\vec{k})=\frac{1}{\gamma}\frac{1}{L^3}\frac{1}{n_Z}\sum_{p=1}^{n_Z}
\sum_{\vec{y}}\: \widetilde{G}_{\chi_{q'}}^p(y,x_0,\vec{0})^*
\cdot{\rm tr}\left[ \Omega^\dag(y) S(\Lambda) \widetilde{G}_H^p(y,x_0,\vec{k})  \right]
\:e^{-i\vec{k}\cdot\vec{y}}. \label{eq:B_RW_2pt}
\end{equation}
In (\ref{eq:B_RW_2pt}), ``$\mathrm{tr}$'' denotes a trace over spinor indices only.

We also compute correlators with gauge-invariant Gaussian  smearing for the heavy quark
at the source and/or sink. The smearing is performed via the operator
\begin{equation}
 \left(1 + \frac{\sigma}{n_S}\Delta^{(2)}\right)^{n_S}, \label{eq:smear_op}
\end{equation}
where $\Delta^{(2)}$ is a covariant lattice Laplacian and $\sigma$, $n_S$ are the
smearing parameters. The operator (\ref{eq:smear_op}) is inserted in Eq.~(\ref{eq:B_src})
to the left of $\Omega(x)\:\xi^p(\vec{x})$ (for source smearing) and/or to the
left of $G_{\psi_v}(y,x)$ (for sink smearing). 

\begin{figure}[t!]
\includegraphics[width=0.45\textwidth,height=2in]{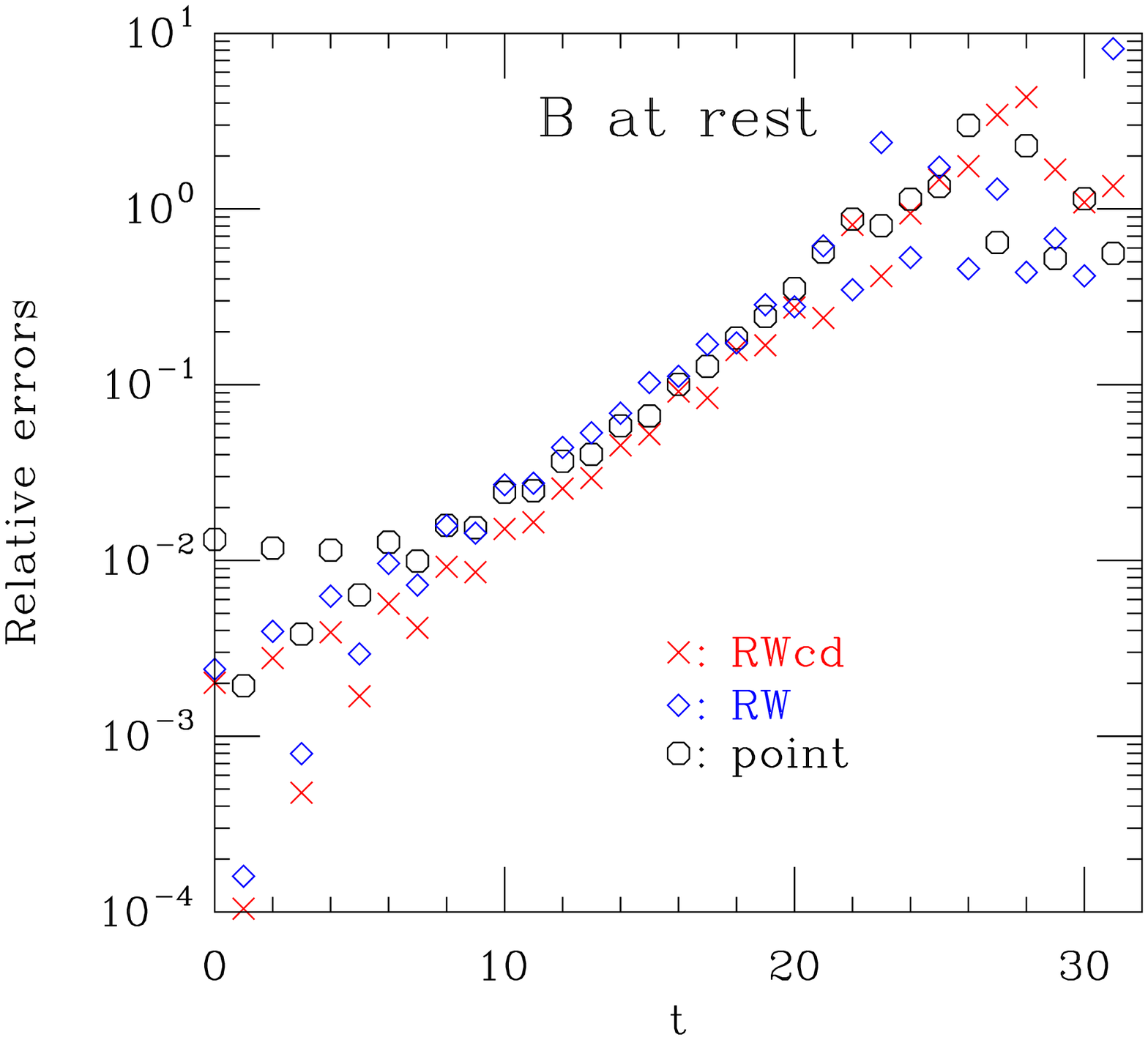}
\hfill
\includegraphics[width=0.45\textwidth,height=2in]{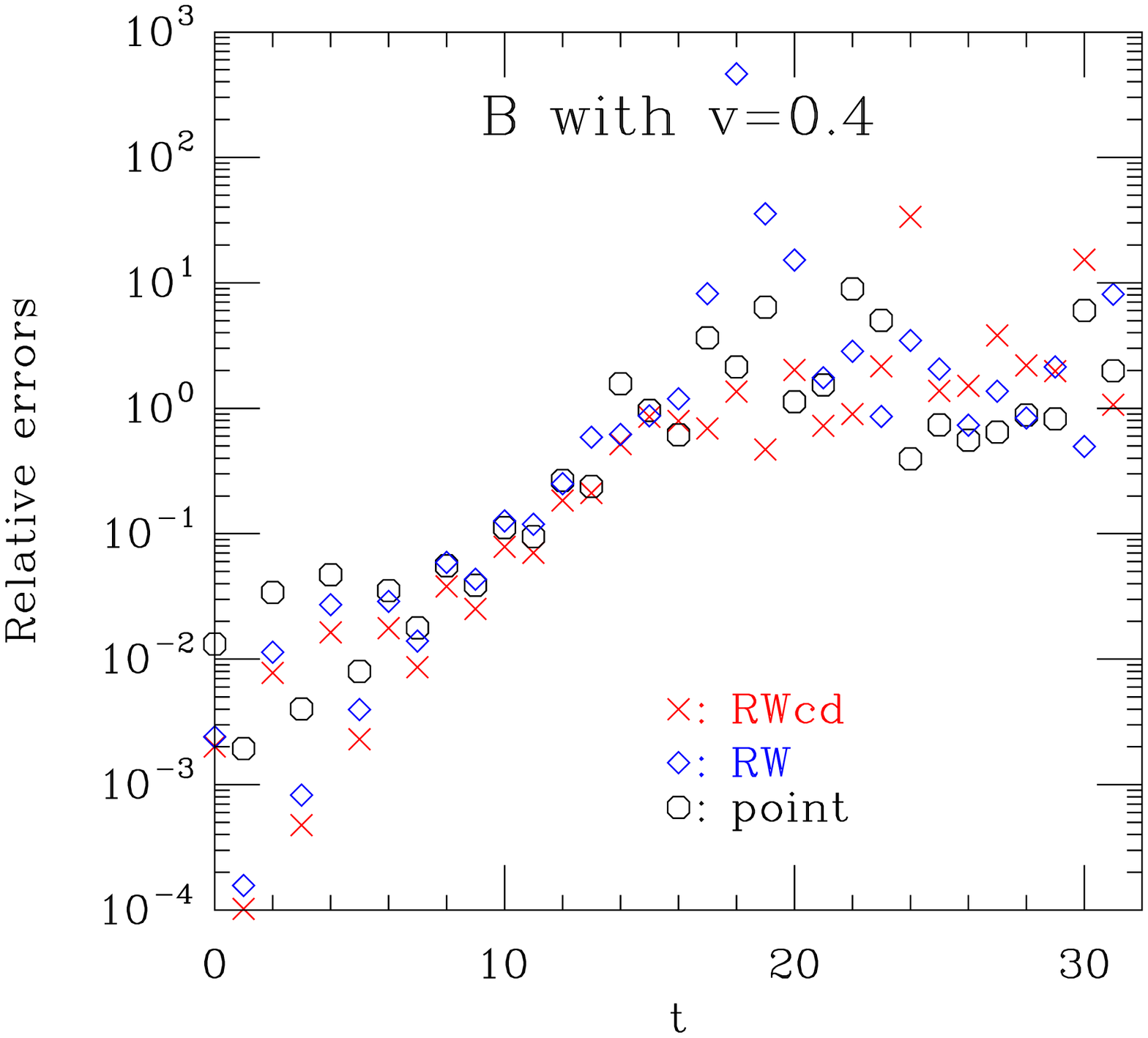}
\caption{Comparison of relative errors of $B$
meson two point functions from the RWcd, RW and point
sources, at $v=0$ (left) and $v=0.4$ (right).}
\label{fig:errB00}
\end{figure}

In Fig.~\ref{fig:errB00} we compare the relative errors of $B$ meson 
two-point functions (without smearing) at $v=0$ and $v=0.4$ from the three sources.
Similar to vector mesons, an advantage of using random wall sources
is only seen at time slices less than 10.

For the heavy-light two-point functions, we use matrix fits with local and smeared sources. The fit function
has the form
\begin{equation}
C_{B}^{s\:s'}(\:t,\vec k)=\sum_{n=0}^{N-1} A_n^{s}\:A_n^{s'*} e^{-E_n t}
+ (-1)^{t+1}\sum_{n=0}^{\widetilde{N}-1} \widetilde{A}_n^{s}\:\widetilde{A}_n^{s'*} e^{-\widetilde{E}_n t},
\end{equation}
where the index $s$ ($s'$) labels the type of smearing at the source (sink). As described in Sec.~\ref{sec:ll2pt},
we actually use the logarithms of the energy splittings and the relative excited state amplitudes as the fit
parameters.

\begin{figure}
\begin{center}
\includegraphics[clip=true,width=0.5\textwidth]{B_v04_2x1_matrix_mom_0_0_0_bay66_t2}
\hfill
\includegraphics[clip=true,width=0.45\textwidth,height=2in]{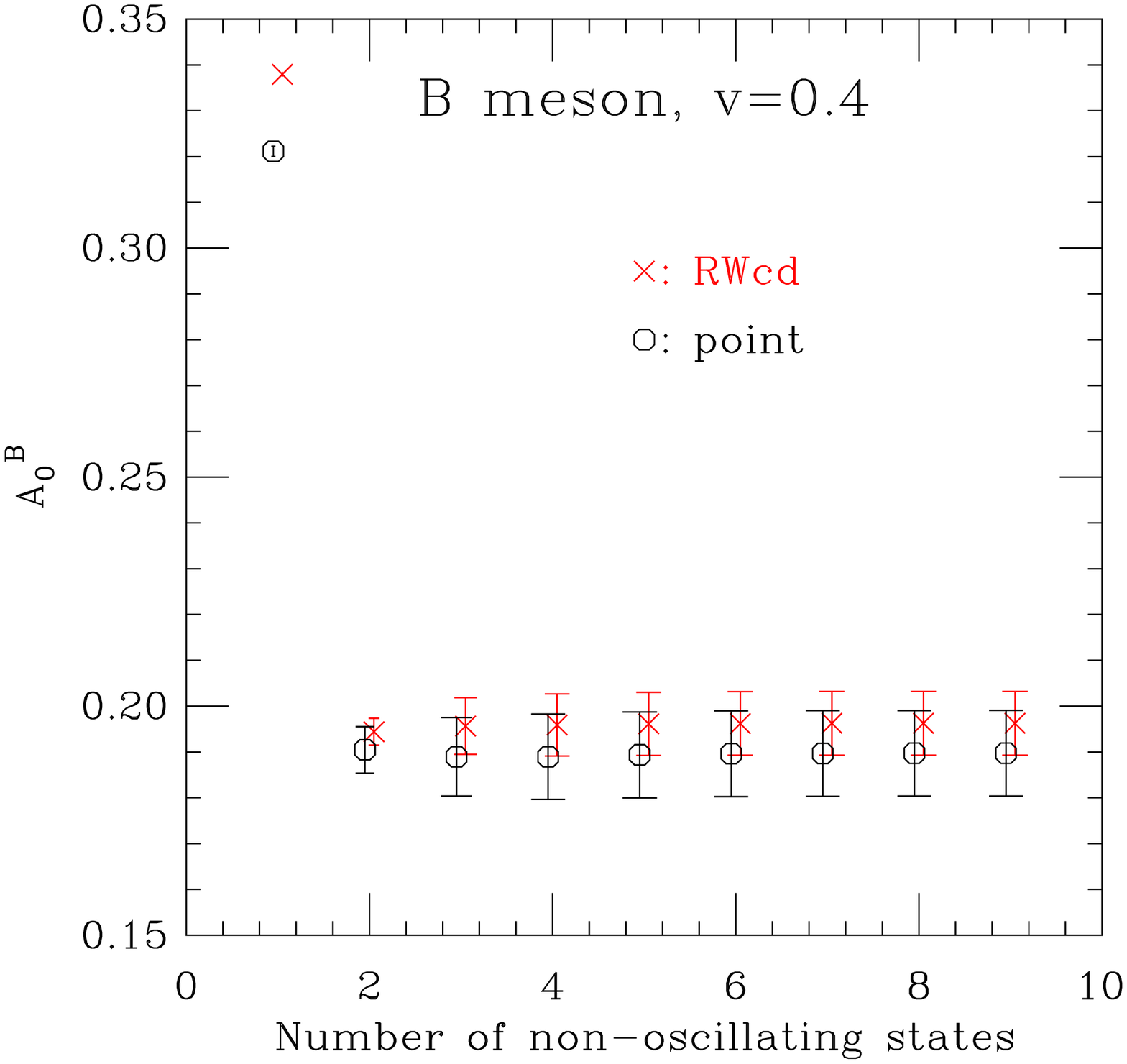}
\end{center}
\caption{\label{fig:fitBv04}
Left: A constrained 6+6 exponential matrix fit for the $B$ meson at $v=0.4$
(RWcd source; fit range $t=2...32$). Right: Fit results for the local $B$ meson amplitude
at $v=0.4$ vs the number of exponentials $N=\widetilde{N}$ in the Bayesian fit, for the RWcd and point sources.}
\end{figure}

Fig.~\ref{fig:fitBv04} shows an example of a $2\times1$ matrix fit of $B$ meson correlators at $v=0.4$ with
$N=\widetilde{N}=6$. In this fit, comparing the RWcd source with the point source, we see an improvement factor of about 1.4 for the ground state energy and local amplitude. Generally the improvement factors we found for $B$ mesons are small (no improvement in some cases).

\subsection{Three-point functions}
\label{sec:hl3pt}

For $x_0>y_0>z_0$, and writing $t=x_0-y_0,\:T=x_0-z_0$, the exact all-to-all three-point
correlator is
\begin{eqnarray}
\nonumber C_{FJB}(t,\:T,\: \vec{k},\:\vec{p'})&=&\frac{1}{\gamma} \frac{1}{L^3}
\sum_{\vec y,\vec z,\vec x} \:e^{-i\vec{p'}\cdot\vec{x}} e^{-i(\vec{k}-\vec{p'})
\cdot\vec{y}}e^{i\vec{k}\cdot\vec{z}}
\:\:\mathrm{Tr}\Bigg[ G_{\chi_q}^\dag(y,x)\:\: F(x)\:\:\Omega^\dag(y) \:\dgamma_5 \\
&&\times\:\:\mathscr{J}\:\left(\begin{array}{cc} G_{\psi_v}(y,\:z) & 0 \\
0 & 0\end{array}\right)\overline{S}(\Lambda) \:\dgamma_5\: \Omega(z)
\:G_{\chi_{q'}}(z,x)\Bigg], \label{eq:threept_all_all}
\end{eqnarray}
where $F(x)=1$ for a pseudoscalar meson in the final state, $F(x)=(-1)^{x_j}\:\dgamma^j$
for a vector meson in the final state, and $\mathscr{J}$ denotes the gamma matrix
/ derivative operator content of the heavy-light current.
We now define the sequential-source heavy-quark random wall propagator, based on the
light spectator quark random wall propagator $\widetilde{G}_{\chi_{q'}}^p(z,x_0,-\vec{p'})$:
\begin{equation}
 \widetilde{G}_H^p(y,z_0,x_0,\vec{k},\vec{p'})=\sum_{\vec{z}}\:\left(\begin{array}{cc}
G_{\psi_v}(y,z) & 0 \\ 0 & 0\end{array}\right)\overline{S}(\Lambda)\:\dgamma_5
\:e^{i\vec{k}\cdot\vec{z}}\:\Omega(z)\:\widetilde{G}_{\chi_{q'}}^p(z,x_0,-\vec{p'}).
\label{eq:seq_src}
\end{equation}
The random-wall three-point correlator for a pseudoscalar meson in the final state is then
\begin{eqnarray}
\nonumber C_{5JB,\:{\rm RW}}(t,\: T,\: \vec{k},\:\vec{p'})&=&\frac{1}{\gamma}\frac{1}{L^3}
\frac{1}{n_Z}\sum_{p=1}^{n_Z}\sum_{\vec{y}}\:\widetilde{G}_{\chi_{q}}^p(y,x_0,\vec{0})^*
\cdot {\rm tr}\left[ \Omega^\dag(y) \:\dgamma_5\:\mathscr{J}
\: \widetilde{G}_H^p(y,z_0,x_0,\vec{k},\vec{p'})  \right]\\
&&\times\:e^{-i(\vec{k}-p')\cdot\vec{y}},
\end{eqnarray}
and for a vector meson in the final state we have
\begin{eqnarray}
\nonumber C_{jJB,\:{\rm RW}}(t,\: T,\: \vec{k},\:\vec{p'})&=&\frac{1}{\gamma}
\frac{1}{L^3}\frac{1}{n_Z}\sum_{p=1}^{n_Z}\sum_{\vec{y}}
\:\widetilde{G}_{\chi_{q}}^p(y,x_0,j)^*
\cdot {\rm tr}\left[ \dgamma^j \:\dgamma_5 \: \Omega^\dag(y) \:\dgamma_5\:\mathscr{J}
\: \widetilde{G}_H^p(y,z_0,x_0,\vec{k},\vec{p'})  \right]\\
&&\times\:e^{-i(\vec{k}-p')\cdot\vec{y}}.
\end{eqnarray}
For smeared three-point functions, we insert the operator (\ref{eq:smear_op}) in the
equation for the sequential-source heavy-quark propagator (\ref{eq:seq_src}),
to the left of $\Omega(z)\:\widetilde{G}_{\chi_{q'}}^p(z,x_0,-\vec{p'})$.

\begin{figure}
\begin{minipage}[t]{.48\linewidth}
\centerline{\includegraphics[width=\linewidth]{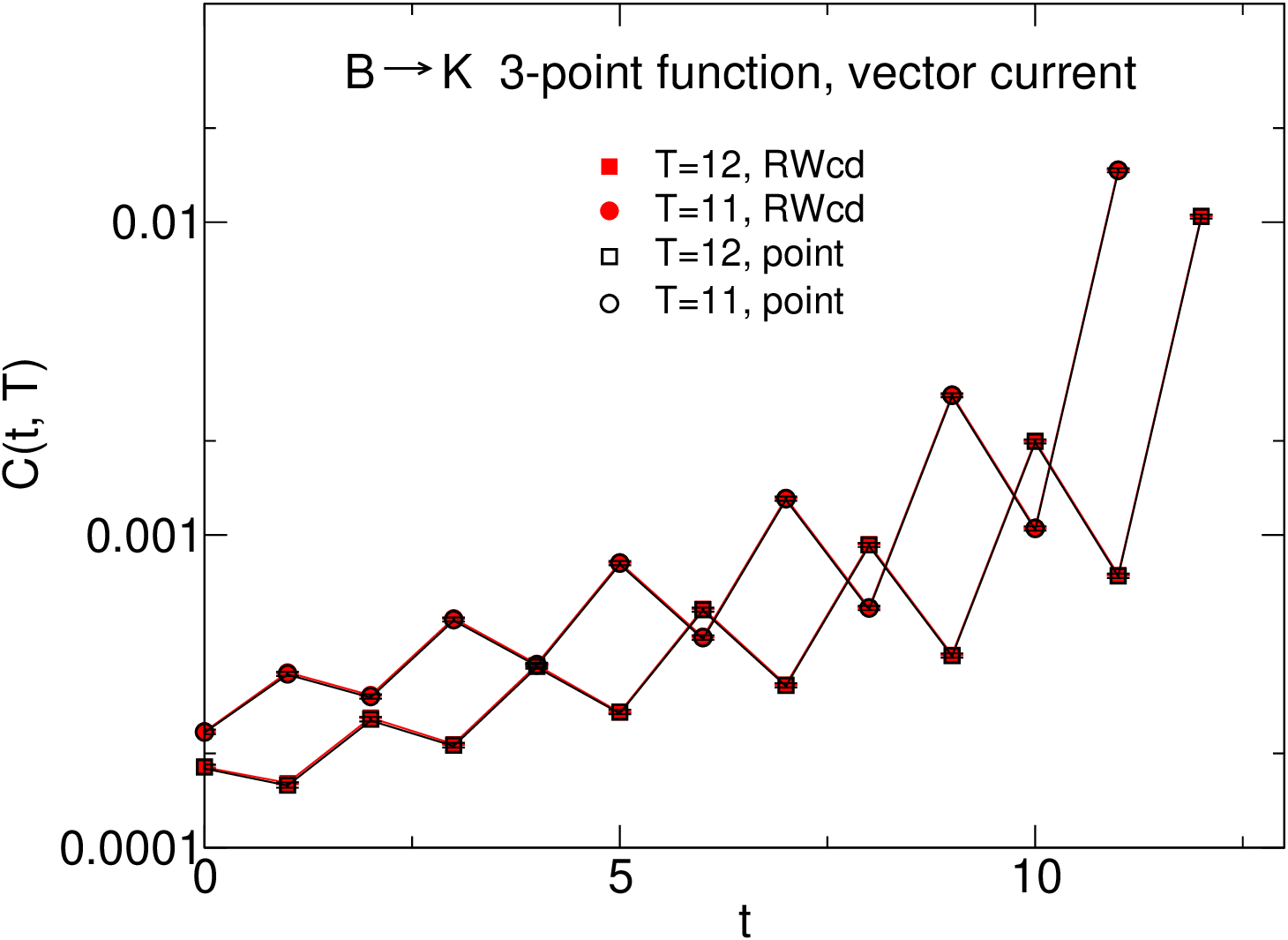}}
\caption{Fits to $B\rightarrow K$ three-point functions at zero recoil (temporal vector current)
from RWcd and point sources (data points coincide).}
\label{fig:fitB2KV0}
\end{minipage}
\hfill
\begin{minipage}[t]{.48\linewidth}
\centerline{\includegraphics[width=\linewidth]{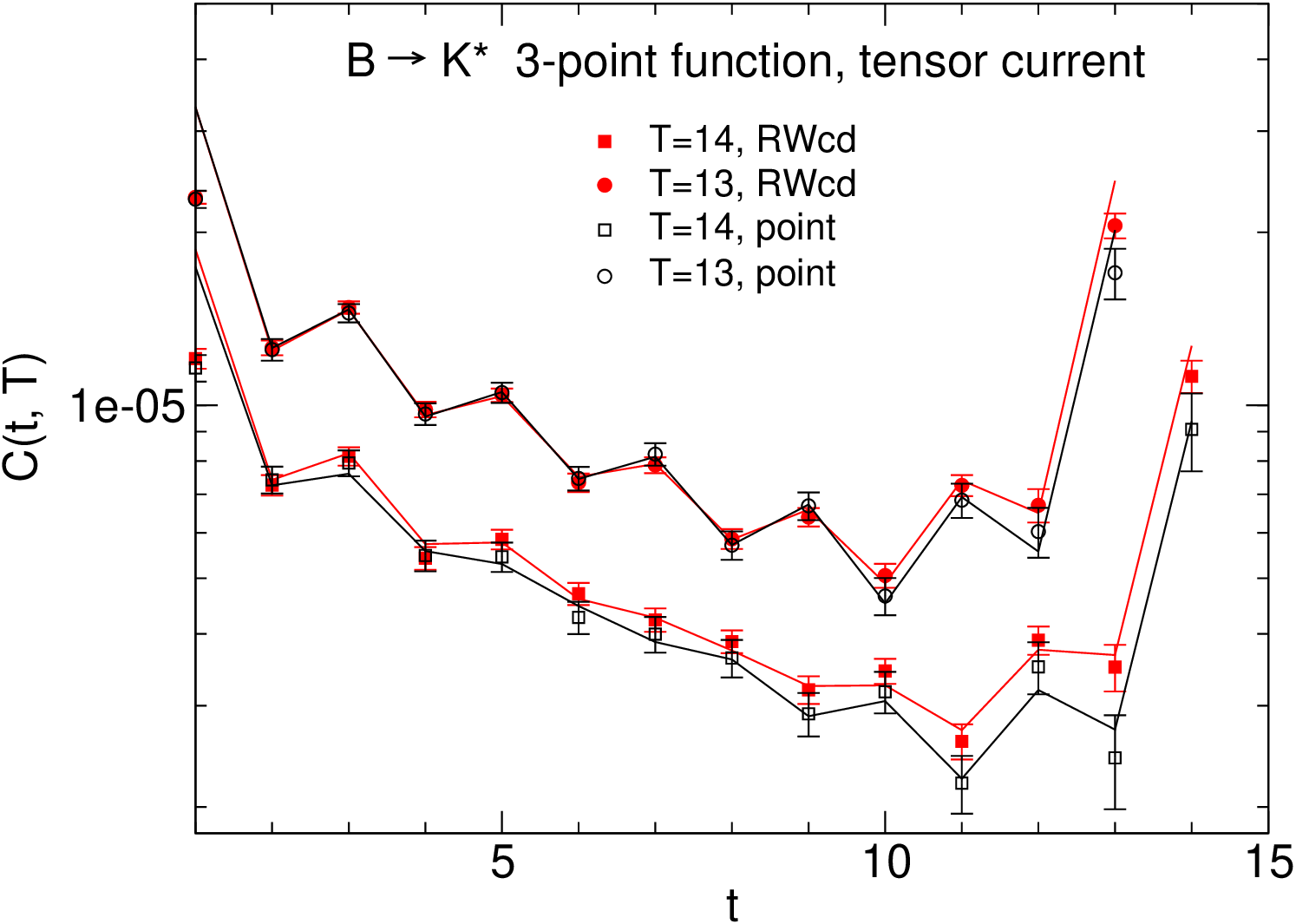}}
\caption{Fits to $B\rightarrow K^*$ three-point functions at zero recoil (current: $\bar s\sigma_{0j}\gamma_5 b$)
from RWcd and point sources.}
\label{fig:fitB2KstarV0}
\end{minipage}
\end{figure}

For the three-point correlators, the fit function has the form
\begin{equation}
C_{FJB}(\vec p',\:\vec k,\:t,\:T)=\sum_{n=0}^{N_F-1}
\sum_{m=0}^{N_B-1}A_{nm}^{(FJB)}e^{-F_n t} e^{-E_m(T-t)}\:\:\:+\:\:\:\rm oscillating\:\:\:terms,
\label{eq:3ptfitfn}
\end{equation}
and again we actually use the logarithms of the energy splittings and the relative excited state amplitudes as the fit
parameters.

Fig.~\ref{fig:fitB2KV0} shows fits to $B\rightarrow K$ three-point functions
at $\vec p' = \vec k = 0$, $v=0$, for the RWcd and point sources. These fits have
$N_B=\widetilde{N}_B=N_K=5$ and $\widetilde{N}_K=0$. Correlators with
$T=11$ and $T=12$ are fitted simultaneously and the range for $t$ is from 0 to $T$.
Comparing the RWcd source to the point source, we find an improvement factor of about 1.6 for
the amplitude $A_{00}$.

Fits of $B\rightarrow K^*$ three-point functions at $\vec p' = \vec k = 0$, $v=0$,
are shown in Fig.~\ref{fig:fitB2KstarV0}. Here, the numbers of exponentials are
$N_B=\widetilde{N}_B=N_{K^*}=\widetilde{N}_{K^*}=6$ and
the fit range is $T=13,14$; $t=2...(T-1)$. A factor of about 1.3 improvement is observed for
the amplitude $A_{00}$ for the RWcd source compared to the point source (but recall that
more inversions were used for the RWcd source).

\section{Simultaneous fits and preliminary form factor results}
\label{sec:FF_results}

The most accurate results for the form factors can be obtained by fitting the two-point
and three-point functions described in sections \ref{sec:ll2pt}, \ref{sec:hl2pt} and \ref{sec:hl3pt}
simultaneously. In the simultaneous fits, the three-point function (\ref{eq:3ptfitfn})
shares the energy parameters $F_n$, $\widetilde{F}_n$ with the light-meson two-point
function, and the energy parameters $E_m$, $\widetilde{E}_m$ with the $B$-meson two-point function.
We fully take into account correlations between all data points.

When multiple values for $T$ in the three-point functions are included, we find that the results for the $B$ meson
energy and two-point amplitude are significantly more accurate compared to fits of the $B$ meson
two-point functions alone. This effect is particularly pronounced for the more precise $B\rightarrow K$
three-point functions.

\begin{figure}
\begin{minipage}[t]{.48\linewidth}
\centerline{\includegraphics[width=\linewidth,height=2in]{B_to_K_f0_f+_point_vs_RWcd_alphaS_tmin10_with_syst_errs}}
\caption{Preliminary results for the form factors $f_0$, $f_+$ for $B\rightarrow K$ decays, obtained from simultaneous non-Bayesian fits
with a wide range of $T$ in the 3-point function. The left-most points have $\vec v=(0.4,0,0)$,
$\vec k=0$ and $\vec p'=2\pi/L\cdot(-1,0,0)$.}
\label{fig:B_to_K_f0_f}
\end{minipage}
\hfill
\begin{minipage}[t]{.48\linewidth}
\centerline{\includegraphics[width=\linewidth,height=2in]{B_to_K_fT_point_vs_RWcd_alphaS_tmin10_with_syst_errs}}
\caption{Preliminary results for the form factor $f_T$ for $B\rightarrow K$ decays, obtained from simultaneous non-Bayesian fits
with a wide range of $T$ in the 3-point function. The left-most points have $\vec v=(0.4,0,0)$, $\vec k=0$ and $\vec p'=2\pi/L\cdot(-1,0,0)$.}
\label{fig:B_to_K_fT}
\end{minipage}
\end{figure}
\begin{figure}[t!]
\centerline{\includegraphics[width=0.48\linewidth]{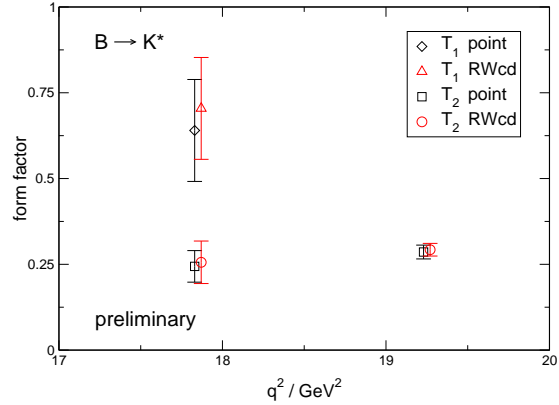}}
\caption{Preliminary results for the form factors $T_1$, $T_2$ for $B\rightarrow K^*$ decays, obtained from simultaneous Bayesian fits with three values for $T$.
The left-most points have $\vec v=0$, $\vec k=0$ and $\vec p'=2\pi/L\cdot(-1,0,0)$.}
\label{fig:B_to_Kstar_T1_T2}
\end{figure}

We have computed the three-point functions for all values of $T$ from 0 to 26 (at $v=0$)
and 0 to 20 (at $v=0.4$), so that we can investigate which range for $T$ gives the best fits.
This investigation is still ongoing. Bayesian fits with the full range $t=0...T$ in the
three-point functions turned out to be much more difficult once more than 3 or 4 values
of $T$ are included (it seems to be impossible to achieve $\chi^2/{\rm dof}\approx 1$ in this case).

Therefore, we have also performed non-Bayesian fits with only $1+1$ exponentials for each meson, including
all available values of $T$ and skipping enough points near the sources so that the contamination from
excited states is seen to be negligible. This only works for $B\rightarrow K$, where the signal is
still good at large times, and even there only at low recoil. Preliminary results for
$f_0$, $f_+$ and $f_T$ obtained with this method are shown in Figs.~\ref{fig:B_to_K_f0_f}
and \ref{fig:B_to_K_fT} respectively. The error estimates are from bootstrap.
At zero recoil, we get $f_0(q^2_{\rm max})=0.869(17)$ from the point source and $f_0(q^2_{\rm max})= 0.889(12)$ from the
RWcd source, an improvement by a factor of about 1.4. However, at $\vec v=(0.4,0,0)$,
$\vec k=0$ and $\vec p'=2\pi/L\cdot(-1,0,0)$ the point source actually gives more accurate results.

For $B\rightarrow K^*$, we used Bayesian fits with three values for $T$, e.g. $T=13,14,15$ and $t=2\:...\:(T-2)$ for $T_1$. The preliminary results for the form factors $T_1$ and $T_2$ are shown in Fig.~\ref{fig:B_to_Kstar_T1_T2}. We have computed the correlation functions for $B_s\rightarrow\phi$ as well, but these still need to be fitted.

All form factor results presented here include the 1-loop radiative corrections in the heavy-light operators
as discussed in Sec.~\ref{sec:current_matching}; a value of $\alpha_s=0.3$ was used here. The $1/m$ corrections
will be included once we have the perturbative results for the mixing coefficients.

\section{Discussion}

\label{sec:discussion}
The stochastic source method we tested reduces statistical errors 
much more effectively in light pseudoscalar mesons than in vector
mesons or heavy-light mesons.
Its effectiveness is further reduced for non-zero momentum, and hence
for lower $q^2$.
For each momentum, and in the case of vector mesons also each polarisation,
additional inversions are needed for the stochastic source method.
Our preliminary results with the random wall sources are generally not as as good as
in \cite{Davies:2007vb}, which may be due to the lower light-quark mass
used here. In calculations with many meson momenta, simply increasing the number
of point sources may be favoured over using stochastic sources
if the total computer time is fixed.

\section*{Acknowledgements}

This work has made use of the resources provided by: the Darwin Supercomputer of the
University of Cambridge High Performance Computing Service (\href{http://www.hpc.cam.ac.uk}{http://www.hpc.cam.ac.uk}),
provided by Dell Inc.\ using Strategic Research Infrastructure Funding from the Higher
Education Funding Council for England; the Edinburgh Compute and Data Facility
(\href{http://www.ecdf.ed.ac.uk}{http://www.ecdf.ed.ac.uk}), which is partially supported by the eDIKT
initiative (\href{http://www.edikt.org.uk}{http://www.edikt.org.uk}); and the Fermilab Lattice Gauge
Theory Computational Facility (\href{http://www.usqcd.org/fnal}{http://www.usqcd.org/fnal}).
We thank the DEISA Consortium (\href{http://www.deisa.eu}{http://www.deisa.eu}), co-funded through
the EU FP6 project RI-031513 and the FP7 project RI-222919,
for support within the DEISA Extreme Computing Initiative.
A.H.\ thanks the U.K.\ Royal Society for financial support. This work
was supported in part by the Sciences and Technology Facilities Council.
The University of Edinburgh is supported in part by the Scottish
Universities Physics Alliance (SUPA).

\end{document}